\documentclass{article}

\usepackage{microtype}
\usepackage{graphicx}
\usepackage{booktabs}
\usepackage{enumitem}

\usepackage{amssymb}
\usepackage{amsmath}
\usepackage{multirow}

\usepackage{algorithm}
\usepackage{algorithmic}

\usepackage{subcaption}

\usepackage{natbib}

\PassOptionsToPackage{hyphens}{url}\usepackage{hyperref}

\usepackage{xr}

\usepackage[accepted]{mlsys2025}

\mlsystitlerunning{DisAgg: Distributed Aggregators for Efficient Secure Aggregation}

\begin{document}

\AddToShipoutPictureBG*{
 \AtPageUpperLeft{
   \hspace*{\paperwidth}
   \raisebox{-59pt}{
     \llap{
       \href{https://www.acm.org/publications/policies/artifact-review-and-badging-current}{
         \includegraphics[height=45pt]{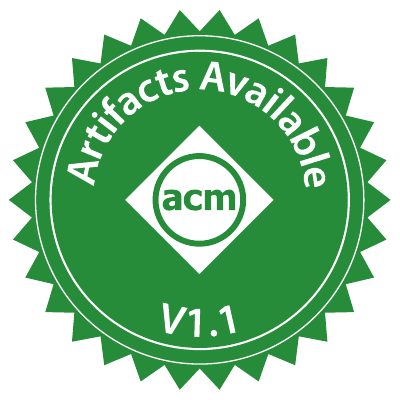}}
       \hspace{1pt}
       \href{https://www.acm.org/publications/policies/artifact-review-and-badging-current}{
         \includegraphics[height=45pt]{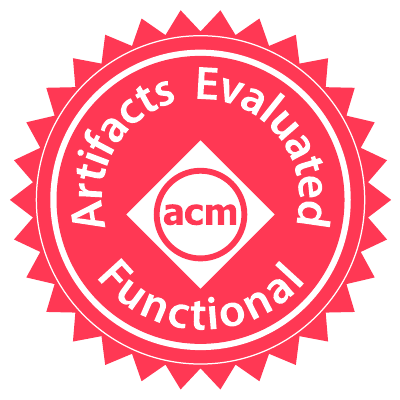}}
       \hspace{1pt}
       \href{https://www.acm.org/publications/policies/artifact-review-and-badging-current}{
         \includegraphics[height=45pt]{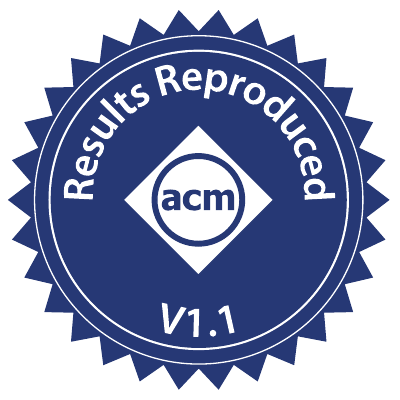}}
       \hspace{80pt}
     }
   }
 }
}

\twocolumn[
\mlsystitle{DisAgg: Distributed Aggregators for Efficient Secure Aggregation in Federated Learning}

\mlsyssetsymbol{equal}{*}

\begin{mlsysauthorlist}
\mlsysauthor{Haaris Mehmood}{sruk}
\mlsysauthor{Giorgos Tatsis}{certh}
\mlsysauthor{Dimitrios Alexopoulos}{certh}
\mlsysauthor{Karthikeyan Saravanan}{sruk}
\mlsysauthor{Jie Xu}{sruk}
\mlsysauthor{Anastasios Drosou}{certh}
\mlsysauthor{Mete Ozay}{sruk}
\end{mlsysauthorlist}

\mlsysaffiliation{sruk}{Samsung R\&D Institute UK (SRUK)}
\mlsysaffiliation{certh}{Information Technologies Institute (CERTH-ITI)}

\mlsyscorrespondingauthor{Haaris Mehmood}{h.mehmood@samsung.com}

\mlsyskeywords{Federated Learning, Secure Aggregation, Privacy, Cryptography}

\vskip 0.3in

\begin{abstract}

Federated learning enables collaborative model training across distributed clients, yet vanilla FL exposes client updates to the central server.  Secure‑aggregation schemes protect privacy against an honest‑but‑curious server, but existing approaches often suffer from many communication rounds, heavy public‑key operations, or difficulty handling client dropouts.  Recent methods like One‑Shot Private Aggregation (\textsc{OPA}) cut rounds to a single server interaction per FL iteration, yet they impose substantial cryptographic and computational overhead on both server and clients. We propose a new protocol called \textsc{DisAgg} that leverages a small committee of clients called \textit{Aggregators} to perform the aggregation itself: each client secret‑shares its update vector to Aggregators, which locally compute partial sums and return only aggregated shares for server‑side reconstruction.  This design eliminates local masking and expensive homomorphic encryption, reducing endpoint computation while preserving privacy against a curious server and a limited fraction of colluding clients.  By leveraging optimal trade-offs between communication and computation costs, \textsc{DisAgg} processes 100k-dimensional update vectors from 100k 5G clients with a 4.6x speedup compared to \textsc{OPA}, the previous best protocol.

\end{abstract}

]

\printAffiliationsAndNotice

\section{Introduction}
Federated Learning (FL) allows many clients to collaboratively train a global model while keeping raw data local~\cite{fedavg,fedopt-survey}. However, client updates can still leak sensitive information through gradient and model inversion attacks~\cite{gradient-inversion,user-level-privacy,deep-leakage}. To mitigate this leakage, several privacy-preserving approaches have been proposed, including differential privacy~\cite{deeplearn-dp,fedlearn-dp,private-learning,dp-fedlearn-client} and homomorphic encryption \cite{privacy-homomorphic}. Among these, Secure Aggregation (SA) is widely adopted for FL, providing strong privacy for individual updates while preserving model utility and maintaining efficiency~\cite{fedlearn-survey}.

In production, federated learning systems must satisfy stringent constraints: intermittent connectivity, client dropouts, adversarial clients, large heterogeneous populations, and tight per‑round latency budgets.  These requirements translate into concrete design goals for SA: low round complexity to mitigate synchronization barriers and straggler effects; robustness to dropouts, corruption, and dynamic participation; minimal per‑client computation and memory for mobile/edge devices; and limited server‑side bottlenecks to sustain high throughput~\cite{fedlearn-survey,flamingo,olympia}.  

Secure Aggregation (\textsc{SecAgg}) \cite{practical-secagg} protects individual contributions by having clients generate pairwise cryptographic keys and add complementary random masks to their model updates; the masks cancel when summed, revealing only the total update while preventing an honest‑but‑curious server from recovering any single client’s data. To tolerate dropouts, missing masks are reconstructed via secret‑sharing seeds from surviving participants\footnote{For details on Shamir's Secret Sharing, see appendix.}. However, these designs require multi‑round exchanges and heavy per‑client cryptographic work---each of the \(N\) clients performs \(O(N)\) key exchanges, so the server must route and validate \(O(N^{2})\) messages \cite{practical-secagg}. Subsequent variants such as \textsc{SecAgg+}, \textsc{FastSecAgg}, \textsc{LightSecAgg}, \textsc{Flamingo}, and \textsc{Turbo‑Aggregate} \cite{secagg-plus, fastsecagg, lightsecagg, flamingo, turbo-aggregate} mitigate some of this cost but the need for multiple rounds and recovery steps remains a scalability bottleneck for dynamic, large‑scale deployments \cite{fedlearn-survey, flamingo}. One‑shot protocols, such as \textsc{OPA}~\cite{opa,willow}, address this by allowing each client to upload a single message per FL iteration, proceeding once a sufficient threshold of client and committee messages is collected; this eliminates multi‑phase handshakes, reduces synchronization barriers, and improves straggler resilience, at the expense of higher per‑round cryptographic overhead that scales with model dimension and committee size.

\begin{figure}[ht]
    \centering    \includegraphics[width=0.85\columnwidth]{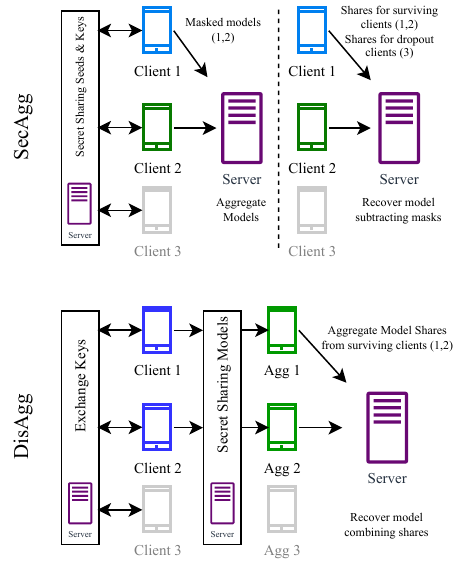}
    \vspace{0.01cm}
    \caption{Overview of the \textsc{SecAgg} protocol compared with the proposed \textsc{DisAgg}. \textsc{SecAgg} uses masks for the models to hide individual inputs and the server aggregates the masked models, whereas \textsc{DisAgg} secret-shares \textit{part of} the model parameters to the Aggregators for them to perform the partial aggregation instead of the server.}
    \label{fig:protocol-overview1}
\end{figure}

We aim to preserve the reduced synchronization advantages of recent protocols while improving computational efficiency at both the client and the server.  To this end, we present \textsc{DisAgg}, a novel SA protocol that distributes aggregation across a small committee of clients (the “Aggregators”). Figure~\ref{fig:protocol-overview1} contrasts \textsc{SecAgg} and \textsc{DisAgg} at a high level. Building on prior committee‑based and distributed‑summation ideas \cite{fastsecagg,lightsecagg}, \textsc{DisAgg} has clients secret‑share their model updates directly to the Aggregator committee.  The Aggregators compute partial sums locally and return only aggregated shares for server‑side reconstruction.  This approach eliminates local masking, reduces regular client computation, and lowers server reconstruction cost, while keeping per‑round interaction low.  Like \textsc{OPA}, \textsc{DisAgg} supports asynchronous participation: clients send a single upload per FL iteration, Aggregators accumulate shares as they arrive, and the server reconstructs once enough aggregated shares are available, avoiding barrier synchronization. The added Aggregator‑side communication and computation is mitigated by selecting a small committee, as analyzed in Section \ref{sec:practical-comparison}.

Our contributions are as follows:
\vspace{-1.0em}
\begin{itemize}
    \item We introduce a novel secure‑aggregation protocol that achieves the same security guarantees as prior methods, yet eliminates the heavy overhead of cryptographic masking and dropout‑recovery mechanisms.
    \vspace{-0.5em}
    \item We extend prior theoretical analyses for secret-share threshold selection to include realistic constraints: considering both dropped out and corrupt clients as random variables and incorporating Byzantine fault-tolerance limits, offering new insights for threshold selection.
    \vspace{-0.5em}
    \item We develop a timing analysis framework incorporating both computation and communication complexities to extensively compare against the state-of-the-art protocol \textsc{OPA} without requiring actual simulations. Under realistic deployment settings and practical configurations of $M=1M$ and $N=1M$, our protocol is expected to achieve an $25$-fold speedup over the state-of-the-art \textsc{OPA} scheme.
\end{itemize}

In summary, \textsc{DisAgg} retains the asynchrony and low round complexity desirable for cross‑device FL while improving computational efficiency at both endpoints.  In the following sections, we formalize the protocol and security properties, analyze its complexity, and validate its performance against state‑of‑the‑art protocols, including \textsc{SecAgg}, \textsc{SecAgg+}, \textsc{LightSecAgg}, and \textsc{OPA}~\cite{practical-secagg,secagg-plus,lightsecagg, opa}.

\section{Background and Related Work}
\label{sec:background-related}

\subsection{Federated Learning}
We consider a cross-device federated learning (FL) setting based on the FedAvg algorithm \cite{fedavg}, executed over $T$ synchronous communication rounds \cite{fedopt-survey}. At iteration $t\in\{1,\dots,T\}$, the central server broadcasts the current global model $w_t\in\mathbb{R}^{M}$ to a randomly sampled subset of clients $\mathcal{U}^{(t)}\subseteq\mathcal{C}$, where $\mathcal{C}$ denotes the full population of registered devices. Each selected client $i\in \mathcal{U}^{(t)}$ performs $E$ local stochastic gradient descent (SGD) steps on its private data $D_i$, thereby minimizing its local objective $F_i(w_t)$, i.e., the expected loss on $D_i$, and obtaining an updated local model $w^{i}_{t,E}$. The client then computes the model delta $x_i = w^{i}_{t,E} - w_t \in \mathbb{R}^{M}$ and sends it back to the server. The server aggregates the received deltas and forms the next global model as $w_{t+1}= w_t + \sum_{i\in \mathcal{U}^{(t)}} x_i$. Various FL variants modify this aggregation step in different ways \cite{fedprox,fedopt}.

\subsection{Motivation for Secure Aggregation}
\label{subsec:foundational}
In vanilla FL, the client updates $x_i$ are transmitted in clear text. Even when the server follows the FL protocol faithfully (i.e., is \emph{honest-but-curious}), it can inspect each $x_i$ and potentially infer sensitive information about the underlying private datasets \cite{model_inv_att_1,model_inv_att_2}. Consequently, protecting individual contributions while still enabling the server to compute the global sum has become a central design requirement for practical FL systems.

Secure aggregation addresses this privacy concern by ensuring that the server learns only the \textbf{aggregate} of all client updates and nothing about any single $x_i$ \cite{practical-secagg}. In a typical construction, each client first uses a deterministic rounding scheme to quantize its floating-point update $x_i\in\mathbb{R}^{M}$ to $\hat{x}_i \in \mathbb{Z}_{p}^{M}$, a vector belonging to a finite field of prime order $p$. The quantized vectors are then masked using cryptographic primitives such as pairwise masks to produce $\tilde{x}_i$ \cite{practical-secagg}.

Based on the agreed-upon protocol, the server can eliminate the masks by aggregating the masked client updates, $S_t = \sum_{i\in \mathcal{U}^{(t)}} \tilde{x}_i \bmod p \in \mathbb{Z}_{p}$. The server then de-quantizes $S_t$ back to $\mathbb{R}^M$ and combines it with the current global model to produce the next iterate $w_{t+1}$. Empirically, we observe that the quantization/de-quantization step incurs negligible degradation in model performance provided the field size is sufficiently large (e.g., $p\ge 2^{32}$), which is a standard choice in secure-aggregation implementations.

When combined with distributed differential privacy \cite{ddp}, secure aggregation provides a strong, provable guarantee that individual user data cannot be reconstructed from the communicated messages.

\vspace{-0.5em}
\subsection{Canonical Protocols}
\textbf{SecAgg \cite{practical-secagg}:}
\textsc{SecAgg} is the first practical secure‑aggregation protocol.  Each client $i$ establishes a pairwise symmetric key $k_{ij}$ with every other client $j$ via an authenticated key‑exchange. From $k_{ij}$, a short seed is derived and expanded with a pseudorandom generator into a high‑dimensional pairwise mask $\mathbf{m}_{ij}\in\mathbb{Z}_p^M$.  Client $i$ adds the sum of its incoming masks ($\mathbf{m}_{ij}$) and subtracts the sum of its outgoing masks ($\mathbf{m}_{ji}$), yielding a masked vector $\tilde{x}_i = \hat{x}_i + \sum_{j<i}\mathbf{m}_{ij} - \sum_{j>i}\mathbf{m}_{ji}$.  Because each mask appears with opposite sign in two clients, all masks cancel in the server’s sum, exposing only $\sum_i \hat{x}_i$. 

The major drawback of \textsc{SecAgg} is the complete pairwise graph, which, during the reconstruction phase, requires the server to request shares of dropped-out clients from all remaining clients. This leads to significant communication and verification overhead for large $N$ \cite{practical-secagg,fedlearn-survey}.

\textbf{SecAgg+ \cite{secagg-plus}:}
\textsc{SecAgg+} improves scalability by replacing the complete pairwise graph with a sparse random graph $G$ of a bounded degree.  Clients exchange keys and share seeds only with their neighbors in $G$, reducing the number of masks and secret‑sharing operations per client.  Consequently, total server communication and verification costs drop from $O(N^2)$ to $O(NlogN)$.

\subsection{Subsequent Protocols}

Subsequent works improve over several facets of computation and communication complexity in SA \cite{lightsecagg, fastsecagg, flamingo}. More recently, works pursue a one‑shot approach for secure aggregation, designing protocols in which each client sends a single message per federated‑learning iteration\footnote{Here we don't take into account public key-exchange which is common to all protocols.}, thereby eliminating the multi‑phase handshakes and synchronization barriers of earlier schemes. We introduce the reader to \textsc{OPA} \cite{opa}, a strong baseline for one-shot aggregation. Additional extensions of secure aggregation are mentioned in the appendix.

\textbf{\textsc{OPA} \cite{opa}:}
One-shot Private Aggregation (\textsc{OPA}) tackles round complexity by allowing each client to upload a single, one‑shot message per FL iteration.  This design eliminates multi‑round communication and enables asynchronous participation: the server recovers a sum as soon as the minimum threshold of client and committee messages are received, avoiding barrier synchronization and reducing sensitivity to stragglers. The scheme relies on a random subset of helper clients as `committee' and cryptographic primitives that support additive homomorphism over masked updates. Clients send masked model updates to the server and share corresponding key shares with the committee, enabling the server to remove the aggregate mask using the combined keys received from the committee. However, the shift towards heavier cryptography--LWR‑based masking and packed Shamir‑style encoding--causes client and server costs to scale with model dimension and committee size.

\begin{figure}[t]
    \centering    \includegraphics[width=\columnwidth]{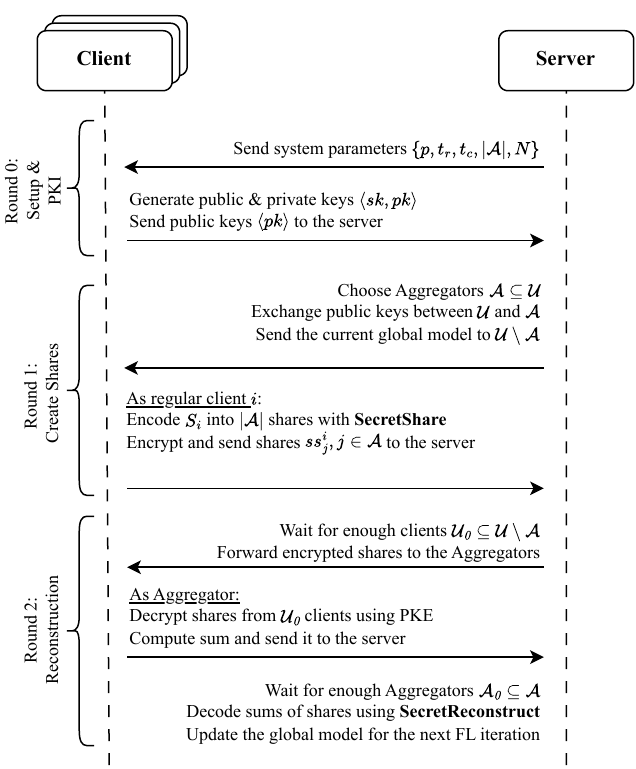}
    \vspace{-0.25em}
    \caption{High-level overview of the proposed secure aggregation protocol \textsc{DisAgg}. $\mathcal{U}$ \& $\mathcal{A}$ denote the set of the clients and Aggregators respectively, $\hat{x}_i$ denotes the secret (model update) for the $i^{th}$ client. PKE stands for the public-key-encryption protocol.}
    \label{fig:protocol-overview}
    \vspace{-2em}
\end{figure}

\subsection{Limitations of Existing Protocols}

Across all lines of work, two patterns emerge. First, canceling-mask designs with multi-phase recovery (e.g., \textsc{SecAgg}/\textsc{SecAgg+}) introduce synchronization barriers and per-party workload that grow with cohort size due to pairwise exchanges, recovery, and checks \cite{practical-secagg,secagg-plus}. Second, minimizing round interaction (e.g., \textsc{OPA}) alleviates synchronization and straggler sensitivity but shifts cost to heavier cryptography and sum recovery that scale with model and committee parameters \cite{opa}.

To overcome such limitations, we introduce a protocol that retain the low‑interaction, one‑shot paradigm of recent works while off‑loading the summation to a small Aggregator committee that operate directly on secret‑shared model updates. To maintain efficiency, our construction leverages standard secret‑sharing primitives and builds on prior distributed‑summation techniques \cite{fastsecagg,lightsecagg,flamingo}. In the following section, we present our protocol \textsc{DisAgg} along with additional notation used throughout the rest of this paper. 

\section{DisAgg}

\vspace{-0.25em}
\subsection{Threat Model}
\vspace{-0.25em}
Following previous works \cite{practical-secagg, lightsecagg}, we adopt an honest-but-curious server threat model that handles up to a fraction $\gamma$ of selected clients colluding at each iteration; additionally our protocol allows for a fraction $\delta$ of selected clients to drop out. Privacy is formally captured as $T$-privacy: any coalition of at most $T$ parties, which includes the server as well as colluding clients, learns no information about honest clients' updates beyond their sum. Leveraging the cryptographic primitives employed in our protocol, information-theoretic security is guaranteed \cite{shamir-secret,lagrange-coding}.

\vspace{-1.0em}
\subsection{Protocol Overview}

The server orchestrates the protocol in three communication rounds per training iteration.  
A round is a complete exchange that starts with a server instruction and ends with the clients’ replies.  
Figure~\ref{fig:protocol-overview} illustrates an overview of our protocol.  
The field prime $p$, reconstruction threshold $t_{\text{r}}$, corruption threshold $t_{\text{c}}$, and the Aggregator group size $A$ are system parameters set by the server.

Unlike canonical schemes \cite{practical-secagg,secagg-plus}, \textsc{DisAgg} delegates the summation to a small subset of clients, called \emph{Aggregators}.  This design is inspired by recent state‑of‑the‑art secure‑aggregation protocols that achieve asynchronous participation with the help of committees \cite{flamingo,opa,willow}.  Section~\ref{sec:aggregators} expands on the role of Aggregators.  For the encryption–decryption messages exchanged between regular clients and Aggregators, we employ a public‑key encryption protocol (Diffie–Hellman) following the approach described in \cite{practical-secagg}.

\textbf{Round 0:} As in most secure‑aggregation schemes \cite{secagg-plus,opa}, the server initially queries clients at random from the entire population, providing them with the system parameters for a given FL iteration.  The round ends once at least $N$ clients have responded with their public keys $pk$, thereby forming the set of participants for the current iteration, $\mathcal{U}$.

\textbf{Round 1:} The server starts this round by randomly selecting $A$ clients from $\mathcal{U}$ to act as Aggregators, denoted $\mathcal{A}$.  It then broadcasts the public keys of the regular clients $\mathcal{U}\setminus\mathcal{A}$ to the Aggregators, and vice versa.  Concurrently, the server transmits the current global model to all regular clients.  Each $i^{th}$ regular client trains the global model on its private data to produce a local model $S_i$, which is then split into secret shares.  The $j$th share, $ss_j^i$, is encrypted with the $j$th Aggregator’s public key and sent to the server.

\textbf{Round 2:} The server waits until enough regular clients $\mathcal{U}_0 \subseteq \mathcal{U}$ have returned their secret shares, then forwards these shares to the Aggregators.  The required size $|\mathcal{U}_0|$ is determined by the maximum dropout tolerance $\delta$, i.e. $|\mathcal{U}_0| \ge \delta N$.  Each Aggregator receives the partial updates from all surviving clients in $\mathcal{U}_0$, decrypts them, and locally aggregates the shares.  The resulting partial sums are forwarded to the server.  Finally, the server reconstructs the aggregated model by combining the partial sums from at least $|\mathcal{A}_0|$ Aggregators.  The choice of $\delta$ and the minimum value of $|\mathcal{A}_0|$ affect the protocol’s security and correctness. This is discussed further in Sections~\ref{sec:aggregators}–\ref{sec:security}.

\subsection{Comparison to Existing Protocols}
\textsc{DisAgg} is the first protocol that applies a committee‑based distributed summation directly to model aggregation. Similar to \textsc{FastSecAgg} \cite{fastsecagg}, \textsc{DisAgg} secret‑shares the raw model parameters while leveraging the dropout‑resilient, Lagrange‑coding technique introduced by \textsc{LightSecAgg} \cite{lightsecagg,lagrange-coding}.  However, unlike \textsc{LightSecAgg}, we do \emph{not} broadcast secret shares to every client in the round; only the Aggregators receive them.  Consequently, the computational load on ordinary clients and on the server is substantially reduced.

Although our protocol introduces additional communication for Aggregators compared to \textsc{LightSecAgg} and \textsc{OPA}, we provide both theoretical analyses and empirical evaluations demonstrating superior performance to existing protocols in terms of overall runtime.  Moreover, we explore practical techniques such as increasing the number of Aggregators and packing plain-text updates to mitigate this overhead while preserving the performance gains.

\begin{algorithm}[t]
\caption{Secret Sharing of \textsc{DisAgg}}
\label{alg:secret-share}
\textbf{\{ss\} = SecretShare($S^i$, $A$, $t_c$, $t_r$, $p$)}\\
\textbf{Input}: $S^i$: secret vector of length M (client $i$'s update vector), $A$: number of shares required, $t_c$: corruption threshold, $t_r$: reconstruction threshold, $p$: field prime\\
\textbf{Output}: List of $A$ shares
\begin{algorithmic}[1]
\STATE Reshape the secret vector $S^i$ into a matrix with $L$ rows, each of size $\rho =t_r - t_c$. Let $L = \lceil M / \rho \rceil$. Add random padding if necessary.
\STATE The result is a list of column vectors $\{s^i\} \in \mathbb{Z}_p^L$
\STATE Append $t_c$ additional random vectors $z^i \in \mathbb{Z}_p^L$
\STATE Form the final list $\{y_k^i\}_{k \in [t_r]} = \{s^i, z^i\}$
\STATE Construct a polynomial $f(x) \in \mathbb{Z}_p^L$ via Lagrange interpolation:
\vspace{-0.5cm}
\begin{align}
f(x) &= \sum_{k=0}^{t_r - 1} y_k^i  \cdot L_k(x), \\
\text{where} \quad L_k(x) &= \prod_{\substack{m=0\\m \ne k}}^{t_r - 1} \frac{x - a_m}{a_k - a_m}
\end{align}
\vspace{-0.5cm}
\STATE Let $a$ be the evaluation points for the secrets
\STATE Evaluate $f(x)$ at positions $\{\beta_j\}$ to get shares $\{ss_j^i\} \in \mathbb{Z}_p^L$:
\vspace{-0.2cm}
\begin{equation}
ss_j^i = f(\beta_j), \quad j = 0,1,\ldots,A-1
\end{equation}
\vspace{-0.6cm}
\STATE Let $\beta$ be the evaluation points for the shares
\STATE \textbf{Return} list of shares with corresponding positions: $\{(\beta_j, ss_j^i): \forall j \in [A]\}$
\end{algorithmic}
\end{algorithm}
\vspace{-0.1cm}

\begin{algorithm}[th]
\caption{Secret Reconstruction of \textsc{DisAgg} \vspace{0.0053cm}}
\label{alg:secret-reconstruct}
\textbf{$S$ = SecretReconstruct($\{(\beta_j, ss_j : \forall j \in [\mathcal{A}_1])\}$, $t_c$, $t_r$, $p$)}\\
\textbf{Input}: Shares of the sum of secrets, corruption threshold $t_c$, reconstruction threshold $t_r$, field prime $p$\\
\textbf{Output}: \noindent Reconstructed sum $S$
\begin{algorithmic}[1]
\IF{fewer than $t_r$ shares are provided}
    \STATE \textbf{Return} Error
\ENDIF
\STATE Keep only the first $t_r$ shares
\STATE Construct a polynomial $f(x) \in \mathbb{Z}_p^L$ using Lagrange interpolation:
\vspace{-0.5cm}
\begin{align}
f(x) &= \sum_{k=0}^{t_r - 1} ss_k \cdot L_k(x), \\
\text{where} \quad L_k(x) &= \prod_{\substack{m=0\\m \ne k}}^{t_r - 1} \frac{x - \beta_m}{\beta_k - \beta_m}
\end{align}
\vspace{-0.5cm}
\STATE Let $\beta$ be the evaluation points for the shares
\STATE Evaluate $f(x)$ at the positions for the secrets:
\vspace{-0.2cm}
\begin{equation}
y_j = f(a_j), \quad j = 0,1,\ldots,t_r - 1
\end{equation}
\vspace{-0.6cm}
\STATE Let $a$ be the evaluation points for the secrets
\STATE Recover the list of vectors $\{y_k^i\}_{k \in [t_r]}$ and keep only the first $t_r - t_c$ of them
\STATE Reshape the vectors and remove any padding to recover the expected sum vector $S$
\STATE \textbf{Return} $S$
\end{algorithmic}
\end{algorithm}

\subsection{Secret Sharing}
\label{sec:secret-sharing}

A desirable property for the secret sharing primitive is the support for additive homomorphism between shares and secrets: the sum of the shares should reconstruct the sum of the underlying secrets \cite{fastsecagg, lightsecagg, opa}. To this end, we employ Lagrange Coded Computing (LCC) \cite{lagrange-coding}, which provides this property efficiently. Similarly, \textsc{OPA} \cite{opa} uses packed Shamir Secret Sharing for sharing secret keys \cite{secure-computation-complexity}. While theoretically both methods have similar security guarantees and time complexities, an in-depth comparison between the two is left as future work.

Algorithm~\ref{alg:secret-share} describes the secret sharing process where each client $i$ encodes its update vector $\hat{x}_i = S^i \in \mathbb{Z}_p^M$ and produces ${\{ss_j^i : \forall j\in [A]}\}$ using Langrange interpolation. Next, each Aggregator $j$ receives secrets ${\{ss_j^i : \forall i\in [U_1]}\}$ from surviving clients to produce a share of the `sum of secrets' $ss_j = \sum_{i=0}^{C^t}{ss_j^i}$. Algorithm~\ref{alg:secret-reconstruct} describes the corresponding reconstruction process on the server side. Here, the server concatenates shares of `sum of secrets' $\{ss_j: \forall j \in [|\mathcal{A}_1|] \}$ from the surviving Aggregators $\mathcal{A}_1 \subseteq A$ and decodes it to produce the sum of updates $S$.

\subsection{Aggregators}
\label{sec:aggregators}
Our protocol designates a subset of clients, the \textit{Aggregators}, to compute the sum of clients' secret shares. Their correctness and security, which will be formally defined in Sections~\ref{sec:correctness} and \ref{sec:security}, depend on suitably chosen parameters as in
\textsc{SecAgg+} \cite{secagg-plus}: $\gamma$ (tolerated fraction of malicious clients), $\delta$ (tolerated fraction of dropped out clients), $P_c$ (probability of security breach) and $P_s$ (probability of reconstruction failure) in each FL iteration. These parameters define the number of Aggregators $A$ (equivalently \textsc{SecAgg+}'s neighbourhood size), the threshold shares $t_r<A$ required by the server to reconstruct the aggregate secret successfully, the corruption threshold $t_c$ of tolerated malicious clients, and the number of secrets $\rho = t_r - t_c$ that can be packed into a single share.

In every FL iteration, the server, assumed to be honest, selects Aggregators at random from the set of the $C_t$ selected clients. An Aggregator is otherwise a regular client who contributes its own model updates, but if chosen as an Aggregator, it instead sums model shares coming from the rest of the clients. 

For enhanced security against a potentially malicious server, we require a trusted source of randomness, such as a random beacon \cite{flamingo}. This means that all parties (clients and the server), once at the beginning of the protocol, obtain a globally shared random seed that will be used for the selection of the Aggregators.

To determine the parameters $A$, $t_r$, $t_c$, and $\rho$, three conditions must be satisfied: (1) the probability that the number of corrupted clients exceeds the corruption threshold $t_c$ must be sufficiently low; (2) the probability that the number of surviving Aggregators falls below the reconstruction threshold $t_r$ must also be sufficiently low; and (3) at least one secret must be embedded in each share. The distribution of corrupted or surviving clients in a sample drawn from a population without replacement follows, by definition, a hypergeometric distribution \cite{hypergeom}. We bound the proportion of corrupted and surviving Aggregators with probabilities $P_c$ and $P_s$, respectively, where $P_c = 2^{-\kappa_c}$ and $P_s = 2^{-\kappa_s}$, with $\kappa_c$, $\kappa_s$ as the security parameters.

Let $X_c \sim \text{Hypergeometric}(N, \gamma N, A)$ and $X_s \sim \text{Hypergeometric}(N, (1 - \delta) N, A)$ be the random variables for corrupted and surviving Aggregators, respectively. As a worst-case scenario, we assume that only honest Aggregators drop out. Then, the following constraints arise from the three conditions that must be satisfied:
\begin{equation}
\Pr(X_c \ge t_c) \;=\; 1 - \mathrm{cdf}_{X_c}(t_c - 1) \;<\; P_c
\label{eq:cdf_corrupted}
\end{equation}
\begin{equation}
\Pr(X_s < t_r) \;=\; \mathrm{cdf}_{X_s}(t_r - 1) \;<\; P_s
\label{eq:cdf_surviving}
\vspace{-0.2cm}
\end{equation}
\vspace{-0.2cm}
\begin{equation}
\ t_c \;<\; t_r 
\label{eq:thresholds}
\vspace{-0.2cm}
\end{equation}

where $cdf$ denotes the cumulative distribution function of the random variables, and can be expressed as $cdf_X(t) = Pr(X \le t)$. Using an iterative search algorithm, we can fix the desired number of packed secrets $\rho \ge 1$ and determine the values of $A$, $t_c$, and $t_r$ (with $0 < t_c < t_r < A$) that satisfy the above constraints. In the special case of a single secret ($\rho = 1$), we can compute the minimum required number of Aggregators $A$ that ensures both correctness and security.

In order to support Byzantine fault tolerance, there is another constraint, that is, $\gamma_a + \delta_a < 1/3$ \cite{opa, flamingo}, where $\gamma_a, \delta_a$ are the fraction of the corrupt Aggregators and the dropout Aggregators respectively. Both of these random variables follow Hypergeometric distribution. To compute the probability distribution of the sum of these two random variables, we use the direct convolution technique \cite{hypogeom}. Let $X = X_c + X_d$, where $X_c$ the random variable for corrupt Aggregators defined above, and $X_d \sim \text{Hypergeometric}(N, \delta N, A)$ the random variable for dropout Aggregators. The probability mass function for $X$ is defined as, $\textit{pmf}_{X} = \textit{pmf}_{X_c} \ast  \textit{pmf}_{X_d}$, where $\ast$ denotes convolution. The threshold for $X$ is $A/3$, therefore, we must satisfy the condition, 
\vspace{-0.2cm}
\begin{equation}
    Pr[X<A/3] = 1 - \textit{cdf}_{X}(A/3 - 1) < P_c = 2^{-\kappa_c}.
\end{equation}

We use the same security parameter $\kappa_c$ that defines the threshold probability $P_c$, as in Equation~\ref{eq:cdf_corrupted}. The Aggregators size $A$, can be found again by an iterative search algorithm using the above condition. Once we obtain a value for $A$, we use it as a minimum value for Equations~\ref{eq:cdf_corrupted},~\ref{eq:cdf_surviving} to compute the thresholds $t_c$ and $t_r$.

Finally, we must account for the fact that the size of the Aggregator group directly impacts their communication overhead. The data size of all of the model shares that each Aggregator receives from $N$ clients is approximately $\mathcal{O}(QM)$, where $Q = N / \rho$. A higher packing factor $\rho$ implies more Aggregators $A$ (Equations~\ref{eq:cdf_corrupted}--\ref{eq:thresholds}), reducing per-Aggregator communication at the cost of increased total communication and computation across the Aggregator group. In practice, $\rho$ can be tuned to balance these effects, while $A$ is obtained via the aforementioned search algorithm, with a reasonable compromise value for $Q$ typically on the order of 10. 

Under LCC-based secret sharing, each share sent to an Aggregator has size approximately $M / \rho$, implying a total download of $NM$ across all Aggregators. For instance, with 128-bit field elements, $N = 50\text{k}$, $\rho = 1\text{k}$, and $M = 10\text{k}$, each Aggregator downloads $NM/\rho\cdot128$ bits or about 8\,MB. Section~\ref{sec:mobile-deployability} further quantifies the trade-off between Aggregator download cost and speedup and its implications for mobile deployability.

\begin{table*}[ht]
\centering
\caption{A comparison of computational and communication time complexity across different protocols. $N$ is the number of selected clients for training, $A$ is the size of the Aggregators group, $M$ is the model size. \textsc{SecAgg} is reported from \cite{practical-secagg}, \textsc{SecAgg+}/\textsc{LightSecAgg} from \cite{lightsecagg} and \textsc{OPA} from \cite{opa}.}
\footnotesize
\renewcommand{\arraystretch}{1.1}
\setlength{\tabcolsep}{3pt}
\begin{tabular}{|c|c|c|c|c|c|c|c|}
\hline
& & & \textbf{SecAgg} & \textbf{SecAgg+} & \textbf{LightSecAgg} & \textbf{OPA} & \textbf{DisAgg (Ours)} \\
\hline
\multirow{6}{*}{\textbf{Client}} & \multirow{2}{*}{\begin{tabular}[c]{@{}c@{}}\textbf{Regular}\\\textbf{Offline}\end{tabular}} & \textbf{Comp.} & $O(MN{+}N^2)$ & $O(M\log N{+}\log^2N)$ & $O(M\log N)$ & - & - \\
\cline{3-8}
& & \textbf{Comm.} & $O(N)$ & $O(\log N)$ & $O(M)$ & - & - \\
\cline{2-8}
& \multirow{2}{*}{\begin{tabular}[c]{@{}c@{}}\textbf{Regular}\\\textbf{Online}\end{tabular}} & \textbf{Comp.} & $O(MN{+}N^2)$ & $O(M\log N{+}\log^2N)$ & $O(M)$ & $O(\lambda M{+}A)$ & $O(M\log A)$ \\
\cline{3-8}
& & \textbf{Comm.} & $O(M{+}N)$ & $O(M{+}\log N)$ & $O(M{+}N)$ & $O(M{+}A{+}A)$ & $O(M{+}A)$ \\
\cline{2-8}
& \multirow{2}{*}{\begin{tabular}[c]{@{}c@{}}\textbf{Committee /}\\\textbf{Aggregator}\end{tabular}} & \textbf{Comp.} & - & - & - & $O(\lambda N)$ & $O(NM)$ \\
\cline{3-8}
& & \textbf{Comm.} & - & - & - & $O(N{+}N)$ & $O(NM{+}N)$ \\
\hline
\multicolumn{2}{|c|}{\multirow{2}{*}{\textbf{Server}}} & \textbf{Comp.} & $O(MN^2)$ & $O(MN\log N{+}N\log^2N)$ & $O(M\log N)$ & $O(NM{+}A\log A)$ & $O(M\log A)$ \\
\cline{3-8}
\multicolumn{2}{|c|}{} & \textbf{Comm.} & $O(MN{+}N^2)$ & $O(MN{+}N\log N)$ & $O(NM{+}N^2)$ & $O(NM{+}NA{+}\lambda A)$ & $O(NM{+}NA)$ \\
\hline
\end{tabular}
\label{tab:complexity_comparison}
\end{table*}

\subsection{Correctness}
\label{sec:correctness}

Next, we analyze the correctness, that is, the resilience of the protocol to client dropouts.

\textbf{Theorem 1 (Correctness):} Given a fraction $\delta \in (0,1)$ of client dropouts and a random selection of the set $\mathcal{A}$ for the Aggregators, the \texttt{SecretShare} algorithm generates $A=|\mathcal{A}|$ shares from the secrets $\hat{x}_i$, and the server can successfully reconstruct the sum of the secrets using \texttt{SecretReconstruct}, with probability $1 - 2^{-\kappa_d}$.

\textbf{Proof:} The correctness follows from the key idea of encoding each client's input using the Lagrange polynomial interpolation \cite{lagrange-coding}. It suffices to show that as long as the the server gathers at least $t_r$ shares, it can successfully reconstruct $S$. Since each Aggregator presents a single share of the sum of secrets $ss_j$, We require the active Aggregator set $\mathcal{A}_0\subseteq\mathcal{A}$ to satisfy $|\mathcal{A}_0| \geq t_r$. The probability of this event is $1 - 2^{-\kappa_d}$.

\subsection{Security}
\label{sec:security}

Following the privacy definition from \cite{lagrange-coding}, our secret sharing scheme is \textit{information-theoretically secure} and ensures that up to $T$ colluding parties gain no information about the input.
$T$ is the privacy parameter of the system. Formally, for every subset $\mathcal{T} \subseteq [N]$ of at most $T$ colluding parties, the mutual information $\widetilde{X}_\mathcal{T}$ between the encoded data available to the colluding parties and $X$, which is seen as chosen uniformly at random, must satisfy:
\vspace{-0.25em}
\begin{equation}
I(X ; \widetilde{X}_\mathcal{T}) = 0.
\vspace{-0.5em}
\end{equation}

In other words, for any pair of secrets $s,s'$ and every group of clients of size at most $T$, the shares of $s$ and $s'$ restricted to this group are identically distributed \cite{fastsecagg}.

Algorithm~\ref{alg:secret-share} achieves this guarantee by construction. In particular, the shares are generated using Lagrange interpolation over a finite field $\mathbb{Z}_p$ with the addition of $t_c$ uniformly random vectors acting as masking noise. These random vectors ensure that even if up to $t_c$ shares (i.e., those corresponding to colluding Aggregators) are observed, the original secret remains perfectly hidden.

To see this more concretely, we consider the masking process in our scheme. Let $\mathcal{T} \subseteq [\mathcal{A}]$ be a set of up to $t_c$ colluding Aggregators who obtain a subset of the shares $\{\widetilde{X}_\mathcal{T}\}$. Each share is of the form:
\vspace{-0.5em}
\begin{equation}
\widetilde{X}_\mathcal{T} = S U^{(\mathcal{T})}_{\text{top}} + Z U^{(\mathcal{T})}_{\text{bottom}},
\vspace{-0.5em}
\end{equation}
where $S$ is the reshaped secret matrix, $Z$ is the matrix of random padding vectors, and $U^{(\mathcal{T})}_{\text{top}}, U^{(\mathcal{T})}_{\text{bottom}}$ are the rows of the interpolation matrix corresponding to the data and noise components, respectively.

By construction, $U^{(\mathcal{T})}_{\text{bottom}}$ is a submatrix of a Vandermonde matrix and forms a Maximum Distance Separable (MDS) matrix \cite{lagrange-coding}. Therefore, any $t_c \times t_c$ submatrix of $U^{(\mathcal{T})}_{\text{bottom}}$ is invertible, implying that the random padding $Z$ remains uniformly distributed and independent of $S$, which completely masks the coded data $S U^{(\mathcal{T})}_{\text{top}}$. This completes the argument for $T$-privacy \cite{lagrange-coding}.

In practice, our implementation ensures this privacy guarantee with overwhelming probability $1 - 2^{-\kappa_c}$, where $\kappa_c$ is a statistical security parameter dependent on the field size $p$ and the randomness used during masking. This ensures that no information about the secret is leaked with all but negligible probability, even under $t_c$-collusion.

\section{Time Complexity Analysis}
\subsection{Theoretical Comparison}
Table~\ref{tab:complexity_comparison} reports the offline (setup) and online (per‑iteration) costs for three parties—regular client, Aggregator, and server—across \textsc{SecAgg}, \textsc{SecAgg+}, \textsc{LightSecAgg}, \textsc{OPA}, and our \textsc{DisAgg}. Costs are broken down into computation (comp) and communication (comm) for each role. The `Committee/Aggregator’ column denotes the \textsc{OPA} committee or \textsc{DisAgg} Aggregators (absent in the other schemes).

Detailed derivations of the complexities follow the original works~\cite{practical-secagg,secagg-plus,lightsecagg,opa} and apppendix for the case of \textsc{DisAgg}. Compared with \textsc{OPA}, \textsc{DisAgg} removes server and client
overhead, shifting the load to Aggregators, while both remain one‑shot protocols without offline phases.

\subsection{Refining Complexities}
\label{sec:practical-comparison}

\begin{figure*}[ht]
    \centering
    \includegraphics[width=\linewidth]{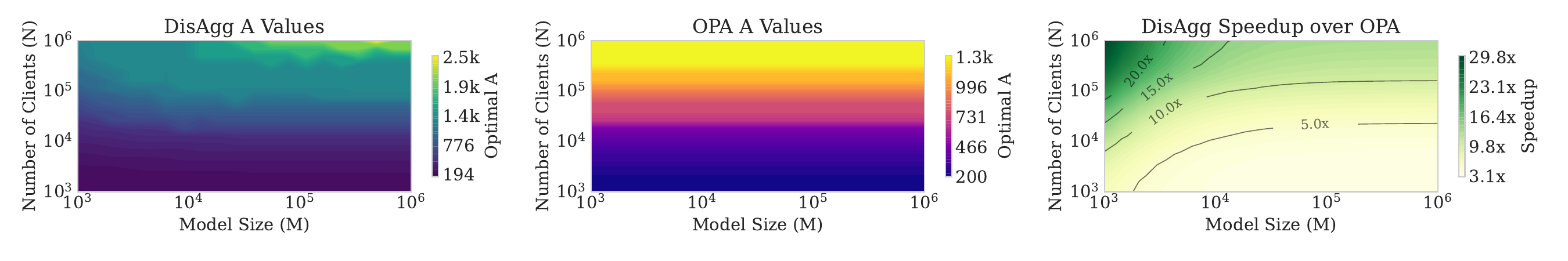}
    \vspace{-1.2cm}
    \caption{Contour plot showing the optimal number of Aggregators $A$ for \textsc{DisAgg} and \textsc{OPA} as well as the expected speedup over \textsc{OPA} across different $(M, N)$ settings under a 5G client connectivity assumption (2 MBps upload / 20 MBps download), with $k=0.3$ and $k_{\text{comp}}=0.66$. Theoretically, \textsc{DisAgg} is faster than \textsc{OPA} for practical cross-device and cross-silo setups: for up to $10^6$ clients per round and models with $10^6$ trainable parameters, \textsc{DisAgg} achieves an estimated speedup of 3.1x -- 29.8x.}
    \label{fig:optimal-a-heatmap}
\end{figure*}

We now refine the asymptotic costs of \textsc{DisAgg} and \textsc{OPA} by inserting the concrete constants that dominate practical performance.  For \textsc{OPA} we expose the packing factor $\rho$ and the security parameter $\lambda$: share generation and reconstruction incur an extra $\mathcal{O}\!\bigl(\tfrac{\lambda}{\rho}\,A\log A\bigr)$ term (packed Shamir secret sharing), while computing Lagrange coefficients adds $\mathcal{O}(A^{2})$ work at both client and server.  The resulting share size is $\mathcal{O}(\lambda/\rho)$ bits, and all $\lambda$‑bit values are transmitted among clients, committee members, and the server.

For \textsc{DisAgg} the dominant overheads are analogous.  Encoding and decoding under Lagrange‑coded computing contribute $\mathcal{O}(A^{2})$ computation at each endpoint, and each model is split into $A$ shares of size $\mathcal{O}(M/\rho)$, where $M$ is the model dimension.  Thus the total communication scales with $M$ rather than $\lambda$.

These refinements reveal complementary trade‑offs. Increasing the packing factor $\rho$, increases the Aggregator set size $A$ as per Section \ref{sec:aggregators} and thus reduces the per‑Aggregator communication in \textsc{DisAgg} but raises the client‑side secret‑sharing and server‑side reconstruction costs; similarly, a larger $\rho$ shrinks \textsc{OPA}’s communication but amplifies the $\mathcal{O}(\lambda/\rho\,A\log A)$ term.  Consequently, an optimal $A$ (or $\rho$) can be chosen given $M$, $N$, $\gamma$, and $\delta$ to minimize overall cost.  A full tabular comparison with all constants is provided in the appendix.

\subsection{Timing Framework}
\label{sec:timing-framework}

We evaluate our protocols under a realistic timing model derived from the theoretical complexities.  Let $k$ denote the total fraction of corrupt and dropped clients, so that $\gamma=1/3\cdot k$ and $\delta=2/3 \cdot k$; we fix $k=0.3$, a common upper bound in secure‑aggregation literature \cite{opa,fastsecagg,flamingo} and within our Byzantine fault‑tolerance limits.  Network parameters are set to the documented Google Cloud egress bandwidth of $25\,\text{GBps}$ for the server \cite{gcp-bandwidth}, and $2\,\text{MBps}$ (upload) / $20\,\text{MBps}$ (download) for clients, reflecting typical 5G links \cite{speed-test}.  The server can serve clients in parallel up to its bandwidth capacity, transmitting model updates in chunks.

Computationally we introduce a constant $k_{\text{comp}}$ that scales the server’s compute cost relative to a client; we use $k_{\text{comp}}=0.66$ for a single CPU core comparison based on standard benchmark ratios \cite{cpu-monkey}. Benchmarks for alternate network settings (4G/3G) are reported in the appendix (Section~\ref{app:alt-network}).

For each $(N,M)$ pair we independently select the number of Aggregators $A$ that minimizes the sum of communication and computation costs for \textsc{DisAgg} and \textsc{OPA}.  The resulting optimal $A$ values are visualized in Figure~\ref{fig:optimal-a-heatmap}, with the right panel showing the corresponding speed‑up of \textsc{DisAgg} over \textsc{OPA}.  A constrained minimum‑Aggregator scenario, mirroring the \textsc{OPA} setup, is examined in Appendix~\ref{app:minimum_a}.

\begin{figure}[ht]
    \centering
    \includegraphics[width=\columnwidth]{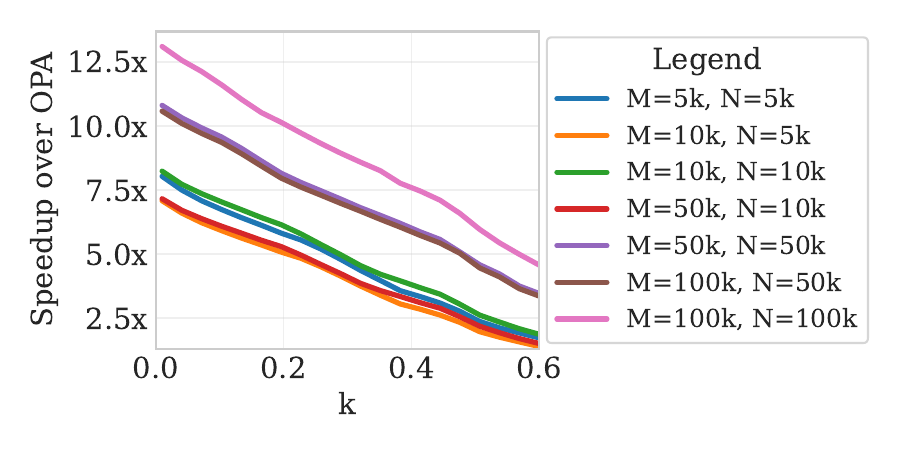}
    \vspace{-1cm}
    \caption{Speedup of \textsc{DisAgg} over \textsc{OPA} as a function of the combined dropout and corruption factor $k = \gamma + \delta$. Results are shown for different $(M, N)$ pairs. The plot demonstrates that \textsc{DisAgg} consistently outperforms \textsc{OPA}, achieving over $4\times$ improvement for practical levels of $k$ up to 0.3.}
    \label{fig:disagg-vs-opa-k}
    \vspace{-0.75em}
\end{figure}

\subsection{Dropout and Collusion Effects}
We analyze sensitivity to combined client instability and adversarial collusion through the aggregate parameter $k= \gamma + \delta$, where $\gamma=1/3\cdot k$ is the collusion fraction and $\delta=2/3 \cdot k$ the dropout fraction. Because both reduce the effective set of honest, active contributors, their impact is largely symmetric in the complexity model (Section~\ref{sec:practical-comparison}). We therefore report relative speedup of \textsc{DisAgg} over \textsc{OPA} across $(M,N)$ configurations while varying $k$. Figure~\ref{fig:disagg-vs-opa-k} shows that even for a practical upper bound of $k=0.3$, \textsc{DisAgg} sustains more than $4\times$ improvement. This indicates resilience of the performance gap under realistic joint instability and collusion assumptions.

\subsection{Deployability on Mobile Devices}
\label{sec:mobile-deployability}

Following the analysis in Section~\ref{sec:aggregators}, we examine \textsc{DisAgg}'s deployability at mobile endpoints, with the primary focus on Aggregator downstream load. Each Aggregator receives shares corresponding to approximately $Q M$ field elements per round, where $Q \!=\! N/\rho$. $M$ is the model size, $N$ the number of clients per iteration, and $\rho$ the secret packing factor.

In our implementation, each share component is encoded as a single field element over a 128-bit prime, consistent with the security requirements of Shamir secret sharing. Assuming 128-bit field elements, the per-Aggregator downstream volume is approximately $Q\,M \cdot 128$ bits.

In the most demanding configuration we evaluate, for $M\!=\!10^6$ and $N\!=\!10^6$ (5G setting as in Section~\ref{sec:timing-framework}) with the committee size chosen to minimize total computation and communication time (yielding $A = 1607$ and $\rho=1331$), this evaluates to over 12\,GB per Aggregator per iteration. In contrast, \textsc{OPA}'s comparable committee download scales as $Q \,\lambda \cdot 128$ bits, which in the same setting (with $\lambda=2048$ and $\rho=1017$) yields $\sim\!32.2$\,MB.

\begin{figure*}[ht]
    \centering
    \includegraphics[width=\linewidth]{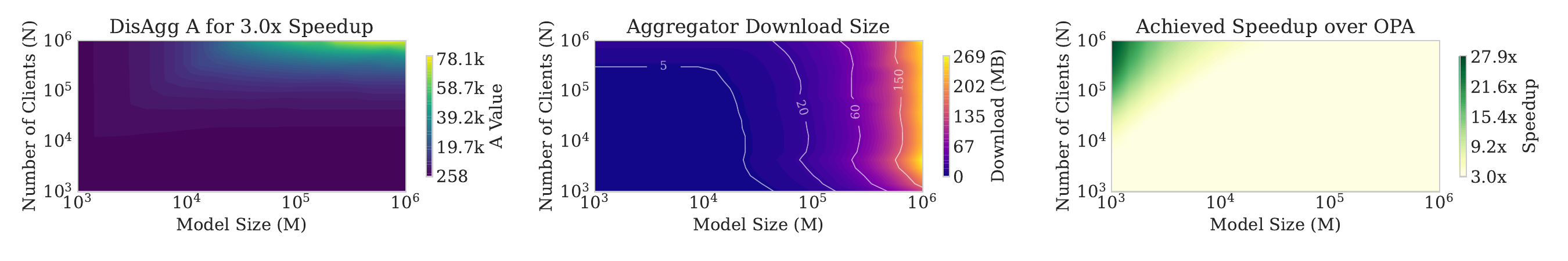}
    \vspace{-1.2cm}
    \caption{Aggregator downstream volume (per iteration) versus committee size, with a target 3$\times$ speedup over \textsc{OPA} under the 5G setting of main paper. Increasing $A$ reduces $Q=N/A$ and thus Aggregator download, trading off against speedup.}
    \label{fig:constant_speedup}
\end{figure*}

There is a tunable trade-off: increasing $\rho$ and thus $A$ decreases $Q$ and hence the Aggregator download, but it can also reduce \textsc{DisAgg}’s speedup over \textsc{OPA}. Figure~\ref{fig:constant_speedup} illustrates this trade-off by enforcing a minimum target speedup equal to the lowest speedup from the 5G analysis (3$\times$ over \textsc{OPA}’s optimal setting) and using a genetic algorithm \cite{storn1997differential} to find the $\rho$ (and corresponding $A$) that balances speedup and minimizes Aggregator download size. The genetic algorithm minimizes a cost function that penalizes configurations falling below the target speedup and, secondarily, larger download sizes.

Specifically, the combined cost is:
\begin{equation}
\text{Cost}
=
10 \cdot 
\underbrace{
\max\!\left\{0,\ \frac{S_{\text{target}} - S_{\text{actual}}}{S_{\text{target}}} \right\}
}_{\text{speedup penalty}}
\;+\;
\underbrace{
\frac{MN/\rho}{MN}
}_{\text{download penalty}}
\end{equation}
where  
$S_{\text{actual}} = \frac{T_{\text{OPA}}}{T_{\text{DisAgg}}}$ is the achieved speedup,  
$S_{\text{target}}$ is the enforced minimum speedup threshold,  
and $\frac{MN/\rho}{MN}$ is the normalized Aggregator download size. In the extreme case $M=N=10^6$, the required download can be reduced to $\sim$269\,MB while maintaining at least a 3$\times$ speedup.

\begin{figure}[t]
    \centering
    \includegraphics[width=\columnwidth]{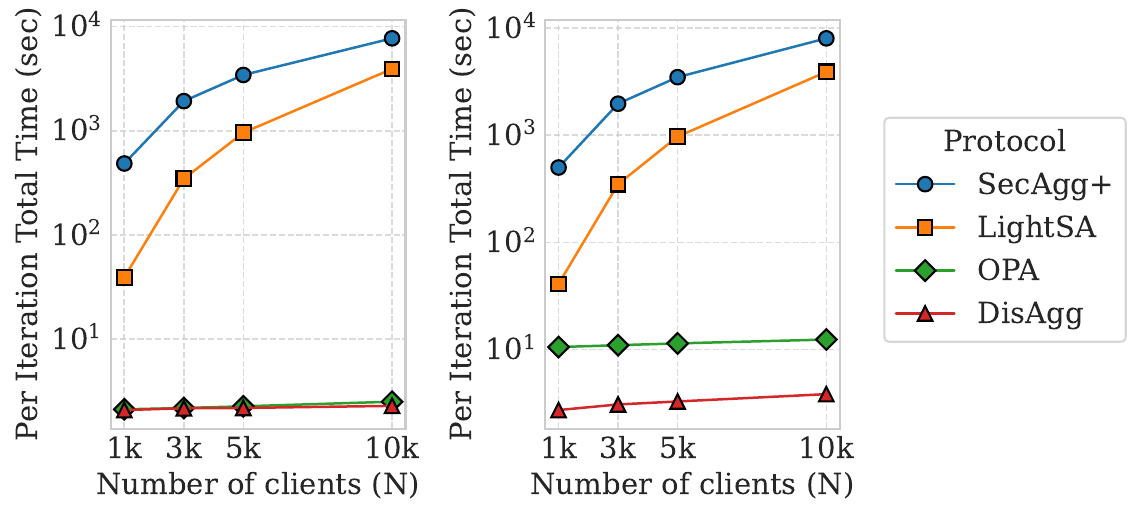}
    \vspace{-0.8cm}
    \caption{Overall combined computation and communication timings per FL iteration for \textsc{SecAgg+}, \textsc{LightSecAgg} (\textsc{LightSA}), \textsc{OPA} and \textsc{DisAgg}. \textsc{DisAgg} is $10\%$ faster for $M=1k$ (left), and 3.2x faster for $M=10k$ (right) over \textsc{OPA}.}
    \label{fig:overall-compare}
    \vspace{-1.0em}
\end{figure}

\section{Experiments}
\label{sec:experiments}

We evaluate \textsc{DisAgg} through four complementary studies: (i) multi-protocol end-to-end timing across established secure aggregation schemes \cite{secagg-plus, lightsecagg, opa}, (ii) a focused large-scale comparison against the one-shot \textsc{OPA} protocol, (iii) sensitivity to combined dropout and corruption, and (iv) impact on multi-round federated model training.

\textbf{Hardware:} All experiments were conducted using server-grade CPUs and a single machine with the same number of parallel processes for all protocols. For training results we make use NVIDIA gpus. The CPU is an Intel(R) Xeon(R) Gold 5220 @ 2.20GHz, with 128GB of RAM, and the GPU is NVIDIA GeForce RTX 2080 Ti, with 12GB VRAM.

\textbf{Simulation Details:} We measure one-iteration timings for finite-field vectors of size $M \in \{1\text{k}, 10\text{k}, 50\text{k}, 100\text{k}\}$ and client cohort sizes $N \in \{1\text{k}, 3\text{k}, 5\text{k}, 10\text{k}, 50\text{k}, 100\text{k}\}$. Reported client and committee values are per-party averages under parallel execution. We developed an in-house simulation framework, incorporating open-source code where available \cite{lightsecagg, practical-secagg}. To obtain practical timings, clients are processed in parallel using 16 processes for $N<50k$ and 30 otherwise. The procedure is divided into discrete computation and communication phases, measured separately: execution time is recorded via a timer for computation, while communication is estimated by translating transferred data (server–client) into time units based on per-client/committee and server bandwidth constraints.

\textbf{Hyperparameter Choices:}
To ensure a fair comparison with \textsc{OPA}, all client updates are quantized to $p - \lceil \log_2 N \rceil$ bits, where $p=53$ is the field size used in \textsc{OPA} after rounding. \textsc{OPA} combines this with ciphertext modulus $q=128$ bits to achieve 129-bit security under the Learning With Errors (LWE) hardness estimator~\cite{lwe-estimator}. Due to rounding errors from LWR-based masking, it reduces plaintext precision by $\lceil \log_2 N \rceil$ bits to ensure correct decryption. We use identical quantization across all protocols to avoid timing or accuracy differences.

Unlike \textsc{OPA}, which relies on the LWR security parameter $\lambda$ for computational masking, \textsc{DisAgg} achieves information-theoretic security through Lagrange Coded Computing (LCC) secret sharing over a subset of Aggregators. The number of Aggregators $A$ is selected using privacy and correctness thresholds $\sigma=40$ and $\eta=40$, providing the same failure and dropout probabilities as prior methods~\cite{secagg-plus,flamingo}. Unless otherwise stated, we use $\gamma=\delta=0.1$ and conduct FL training with $1\text{k}$ clients over 30 iterations.

\begin{figure}[ht]
\centering
\begin{subfigure}{\columnwidth}
    \flushleft
    \vspace{-0.2cm}
    \hspace{0.5cm}
    \includegraphics[width=0.9\columnwidth]{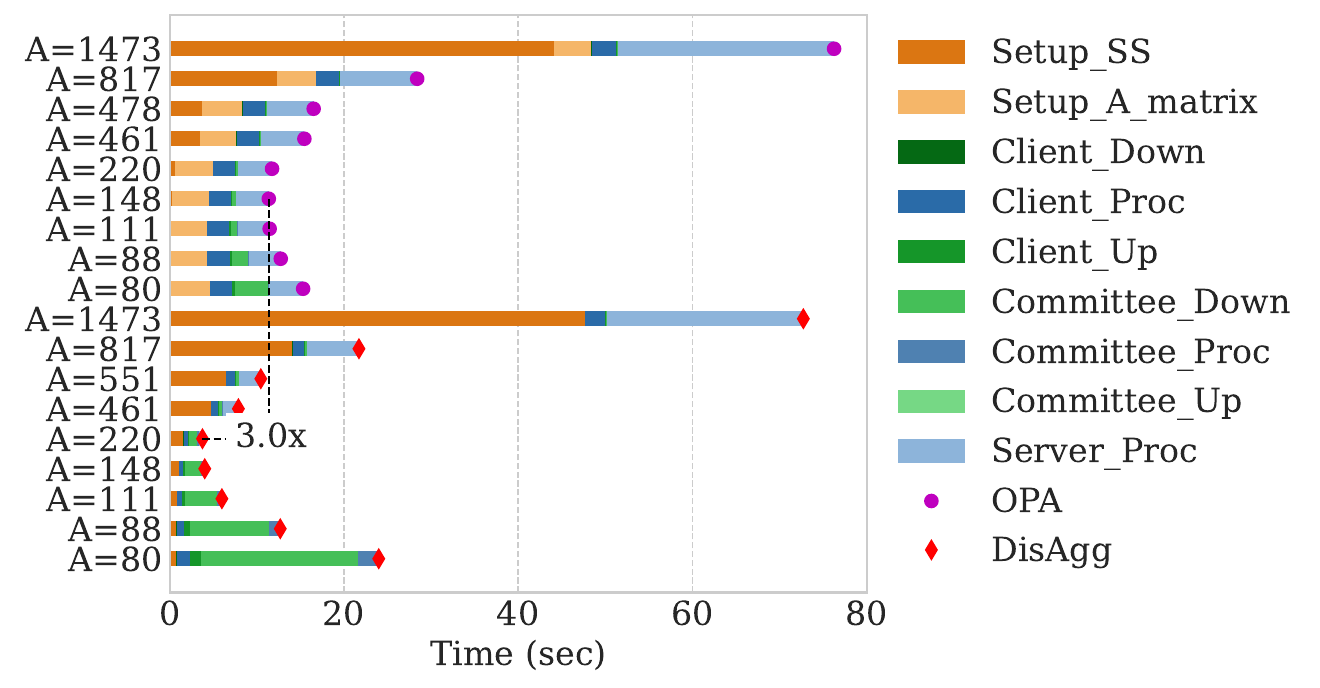}
    \vspace{0.1cm}
\end{subfigure}
\begin{subfigure}{\columnwidth}
    \flushleft
    \hspace{0.5cm}
    \includegraphics[width=0.6\columnwidth]{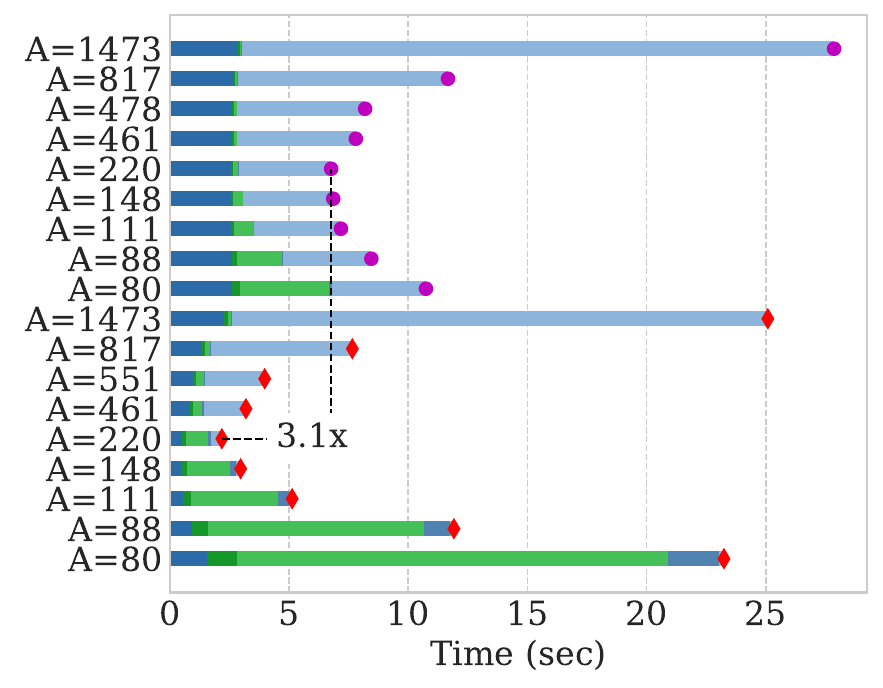}
    \vspace{-0.4cm}
\end{subfigure}
    \vspace{-1.0em}
    \caption{Speedup of \textsc{DisAgg} over \textsc{OPA} for one FL iteration with $M=10k$, $N=10k$, $\gamma=0.1$, $\delta=0.2$, and varying committee size $A$. \textsc{DisAgg} achieves $3\times$ speedup including the setup phase (top) and $3.1\times$ without it (bottom).}
    \label{fig:k-analysis-bar}
    \vspace{-1.0em}
\end{figure}

\subsection{Multi-Protocol Total Time Comparison}

We compare \textsc{DisAgg} against \textsc{SecAgg+}, \textsc{LightSecAgg}, and \textsc{OPA}, each representing a distinct optimization axis: pairwise masking with recovery (\textsc{SecAgg} / \textsc{SecAgg+}), sparse interaction graphs (\textsc{LightSecAgg}), and one-shot participation (\textsc{OPA}). We report both computational and communication contributions for the primary system roles (clients, committee / Aggregators when applicable, and server), and include the setup phase cost (e.g., key exchange, structural graph or matrix initialization) even when it can be cached for persistent clients.

In Figure~\ref{fig:overall-compare} we see that our protocol achieves state-of-art performance compared to all existing protocols. The first ever protocol, \textsc{SecAgg}, was not tested due to its well documented inefficiency \cite{lightsecagg}. As the number of clients increase, the increase in per iteration total time is the least for our protocol and closely follows \textsc{OPA}. Thus \textsc{DisAgg} is highly suited for practical scenarios of aggregating updates from a large selection of clients.

\begin{table*}[ht]
\caption{A comparison of per stage timings for one FL iteration in seconds rounded to 2 decimal places for $N=100k$ (clients) and $M=100k$ (parameters). Timings for committee and clients are an average across all committee and client members respectively. \textsc{DisAgg} achieves an overall speedup of 4.56x over \textsc{OPA}.
}
\centering
\resizebox{0.95\textwidth}{!}{
    \begin{tabular}{ccccccccc}
    \toprule
    \textbf{Method} & \textbf{Setup} & \textbf{Client Comm} & \textbf{Committee Comm} & \textbf{Client Comp} & \textbf{Committee Comp} & \textbf{Server Comp} & \textbf{Total w/out Setup} & \textbf{Total} \\
    \midrule
    \textbf{\textsc{OPA}} & 50.98& 6.93& 1.15& 27.84& 0.17& 211.19& 247.31& 298.30\\
    \textbf{DisAgg} & 14.00& 13.64& 3.74& 6.21& 18.83& 8.89& 51.33& 65.33\\
    \midrule
    \textbf{Improvement} & 3.64& 0.50& 0.31& 4.48& \textless0.01 & 23.74& 4.81& 4.56\\
    \bottomrule
    \end{tabular}
}
\vspace{-0.2em}
\label{tab:opa-vs-disagg-100k}
\end{table*}

\subsection{DisAgg vs OPA}
\label{sec:disagg-vs-opa}
We extend our experiments to realistic cross‑device scales ($M,N>10^{3}$), evaluating $M=10^{4}$ and $N=5\cdot10^{4}$.  In \textsc{OPA} the dominant early‑iteration cost is the initialization of the $A$ matrix over a 128‑bit field with the standard security parameter $\lambda=2048$, incurring an $O(\lambda M)$ overhead that dwarfs all other setup work (sub‑second). Client‑side computation in \textsc{OPA} is likewise dominated by LWR matrix–vector products, which scale as $O(\lambda M)$.  By contrast, \textsc{DisAgg} requires each client only to perform Lagrange‑Coded Computing encoding (matrix construction plus polynomial evaluation), costing $O(M\log A)$ (or $O(MA)$ in a straightforward implementation)—orders of magnitude smaller than $O(\lambda M)$.

The server must decode and aggregate the masked encodings for \textsc{OPA}, whereas \textsc{DisAgg} merely reconstructs the global model from a compact set of aggregated shares, yielding additional efficiency. \textsc{OPA} is only comparatively lighter for committee computation: \textsc{OPA}’s committee operations involve key combination with $O(\lambda N)$ complexity, while \textsc{DisAgg} Aggregators process with complexity $O((NM)/A)$.

Consequently, despite \textsc{OPA}’s lighter committee work, its slower setup, client, and server phases lead to a substantially higher overall runtime.  Table~\ref{tab:opa-vs-disagg-100k} reports detailed timings for $M=100k,\,N=100k$. The trade‑off between \textsc{DisAgg}'s higher communication cost and lower computational burden is governed by the Aggregator count $A$. As described in Section~\ref{sec:aggregators}, the packing factor $\rho$ can be tuned to adjust $A$ for the same security parameters ($\gamma, \delta, N$) and thus an optimal point in the communication-computation trade-off space can be found. This can be leveraged within a simulation environment to further minimize the total execution time.

For this experiment we sweep the values of the packing factor $\rho \in \{25, 50, 100, 250, 500, 1000\}$ for both \textsc{OPA} and \textsc{DisAgg} in order to pick the best performing configuration for both. For \textsc{OPA} the best value was with $A=461$ whereas for \textsc{DisAgg} $A=830$, for which Table~\ref{tab:opa-vs-disagg-100k} reports the timings. The other values are presented in Figure \ref{fig:100k-rho-analysis} in the appendix.

\subsection{Dropout Analyses}
\label{sec:dropout-tolerance}
\begin{figure}[ht]
\centering
    \flushleft
    \vspace{0.5em}
    \hspace{0.5cm}
    \includegraphics[width=\columnwidth]{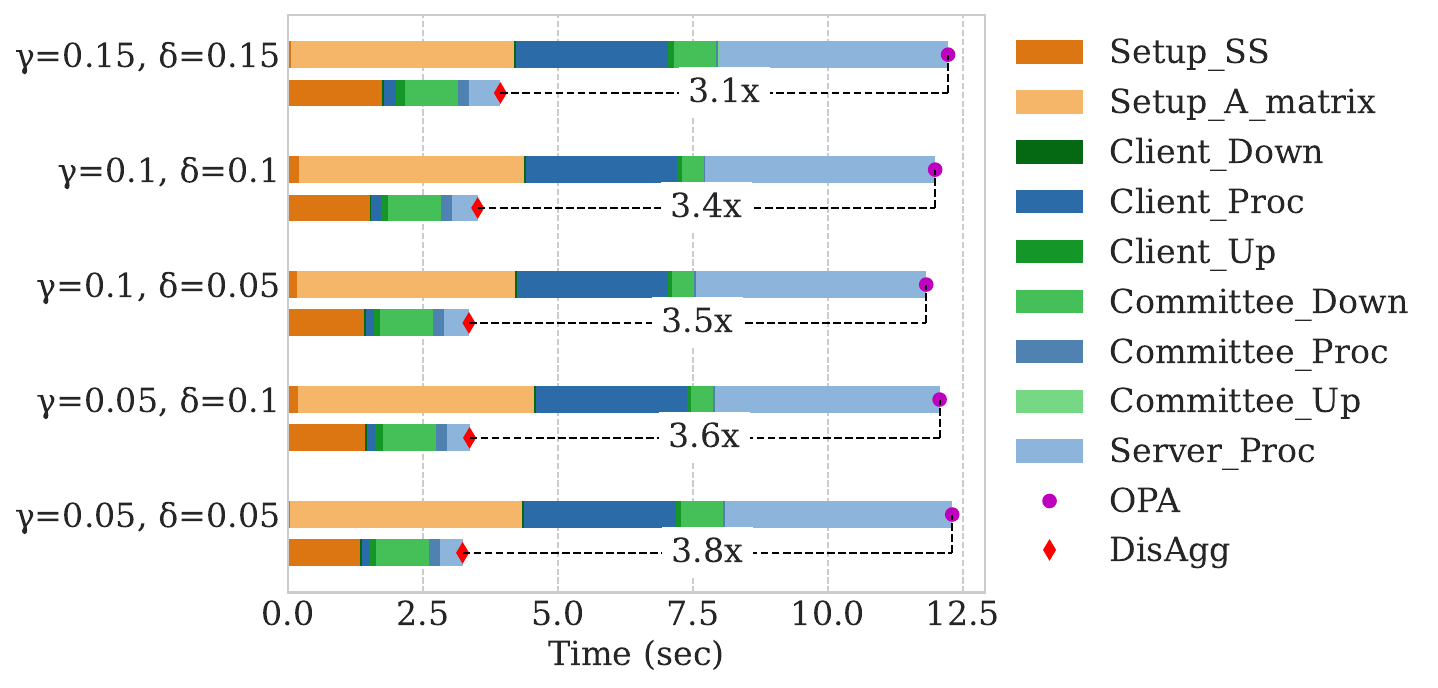}
    \vspace{-1.5em}
    \caption{Speedup of \textsc{DisAgg} over \textsc{OPA} for one FL iteration with $k = \gamma + \delta \text{ given }  \gamma, \delta \in \{0.01, 0.05, 0.10, 0.15\},\, M=10k,\, N=10k$. Top graph depicts timings including setup, whereas bottom graph is without setup phase. }
    \label{fig:dropout-optim-rho}
\end{figure}

We conduct a dropout/collusion sensitivity study with combined instability factor $k = \gamma + \delta$ at values $k \in \{0.01, 0.05, 0.10, 0.15\}$ using a model of size $M = 10k$ parameters and $N = 10k$ participating clients, comparing \textsc{OPA} and \textsc{DisAgg}. Figure \ref{fig:dropout-optim-rho} summarizes the results: for each protocol and pair of $\gamma, \delta$ values, the result with the fastest timing is shown after a grid search on $\rho \in [25, 50, 100, 250, 500]$. \textsc{DisAgg} is robust to client dropout and and collusion with at least 3x speedup over \textsc{OPA} for the same $(\gamma, \delta)$ pair. As expected, for both protocols, the timings are proportional to $k$, as a bigger committee size is required to ensure correctness and security.

\vspace{-0.5em}

\subsection{Aggregator Download Reduction}

As an extension of the discussion in Section~\ref{sec:mobile-deployability} on the speedup versus Aggregator download trade-off, we evaluate this trade-off in an experimental setting using the results derived from the variable-$A$ analysis. From Figure~\ref{fig:k-analysis-bar}, we use the corresponding values of $N$, $M$, and $\rho$ to compute the resulting speedup over \textsc{OPA} and the Aggregator download, following the same calculations described above.

For the setting $N = M = 10k$, Table~\ref{tab:agg_tradeoff_10k} showcases the effect on speedup when increasing $A$: the resulting Aggregator download size in \textsc{DisAgg} can be reduced by more than $2\times$  while still achieving a considerable speedup over \textsc{OPA}.

\begin{table}[h]
\vspace{-1.5em}
\centering
\caption{Speedup and Aggregator download for $N = M = 10k$.}
\vspace{0.25em}
\label{tab:agg_tradeoff_10k}
\begin{tabular}{c c c c}
\hline
\textbf{Speedup over \textsc{OPA}} & \textbf{$A$} & \boldmath$\rho$ & \textbf{Agg. Download} \\
\hline
$3\times$    & $220$ & $100$ & $\sim$16\,MB \\
$1.54\times$ & $461$ & $250$ & $\sim$6.4\,MB \\
$1.18\times$ & $551$ & $305$ & $\sim$5.2\,MB \\
\hline
\end{tabular}
\end{table}

\vspace{-0.5em}
\subsection{Plaintext Recovery}

\begin{figure}[h!]
\centering
\begin{subfigure}{0.48\columnwidth}
    \centering
    \includegraphics[width=\linewidth]{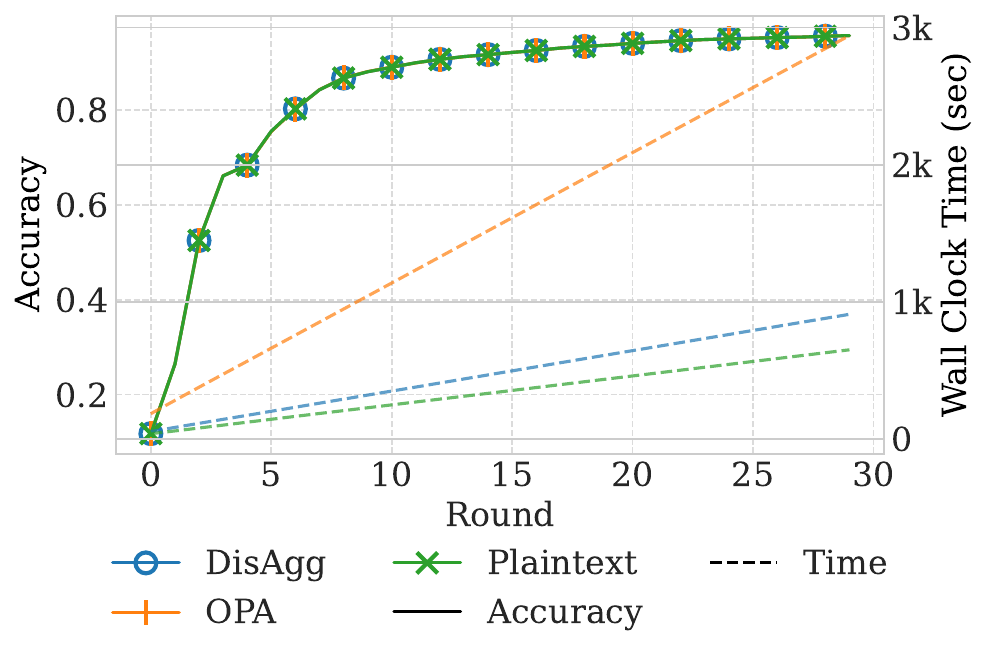}
    \vspace{-1.5em}
    \caption{MNIST}
\end{subfigure}
\begin{subfigure}{0.48\columnwidth}
    \centering
    \includegraphics[width=\linewidth]{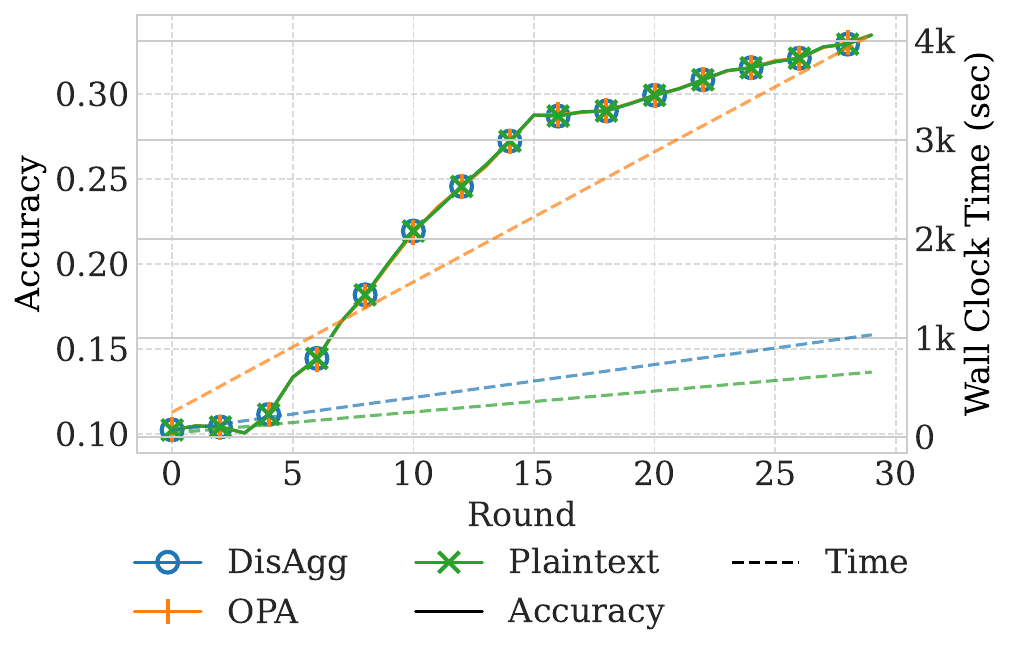}
    \vspace{-1.5em}
    \caption{CIFAR10}
\end{subfigure}
\begin{subfigure}{0.48\columnwidth}
    \vspace{1.0em}
    \centering
    \includegraphics[width=\linewidth]{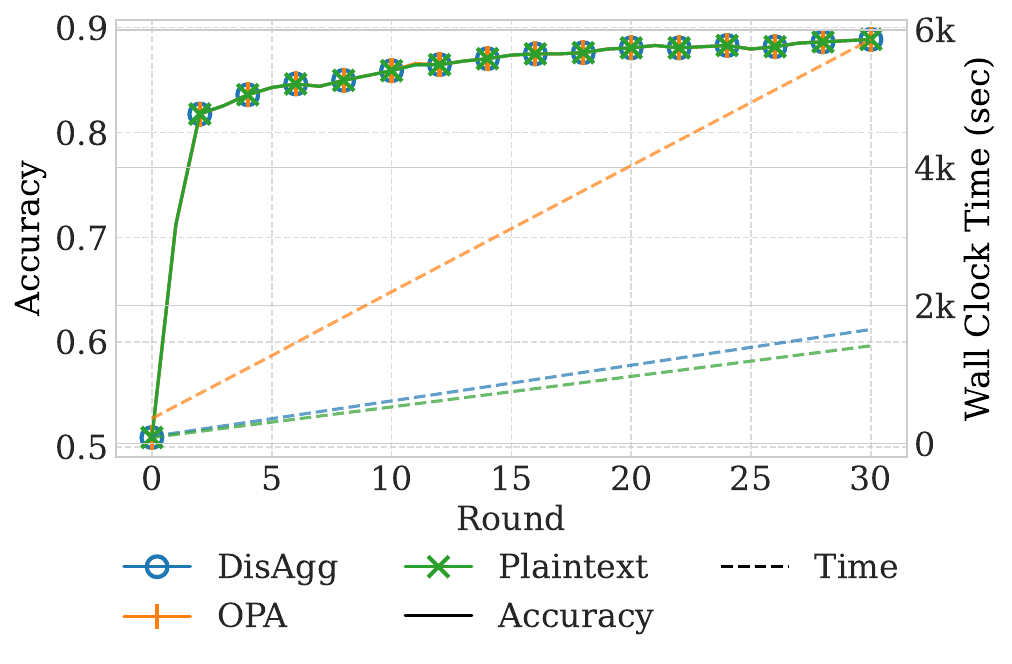}
    \vspace{-1.5em}
    \caption{SST2 (297k)}
\end{subfigure}
\begin{subfigure}{0.48\columnwidth}
    \vspace{1.0em}
    \centering
    \includegraphics[width=\linewidth]{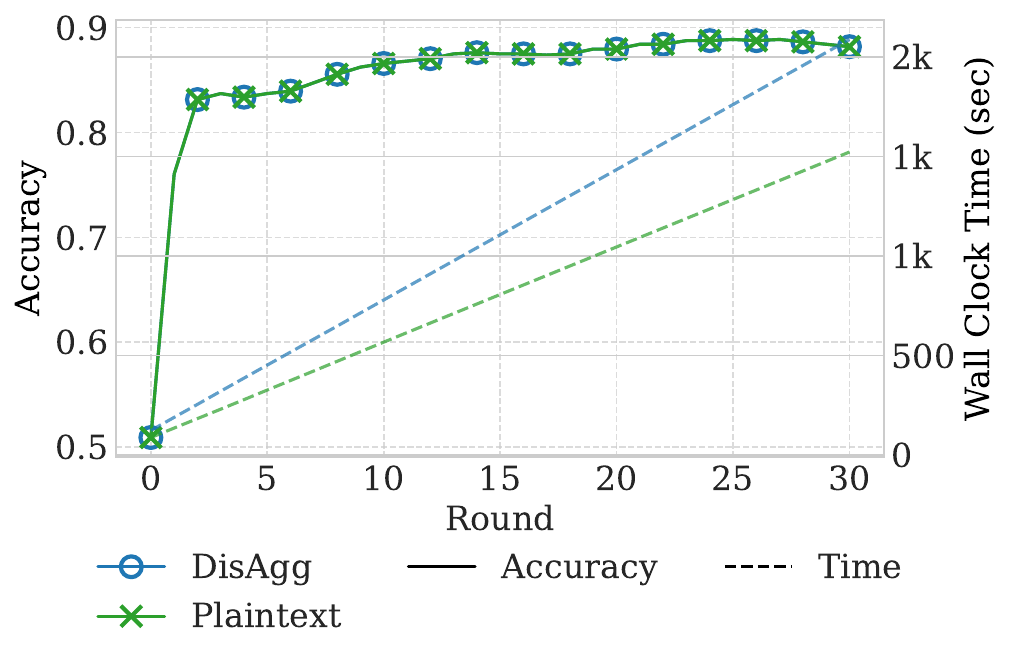}
    \vspace{-1.5em}
    \caption{SST2 (1.1M)}
\end{subfigure}
\begin{subfigure}{0.48\columnwidth}
    \vspace{1.0em}
    \centering
    \includegraphics[width=\linewidth]{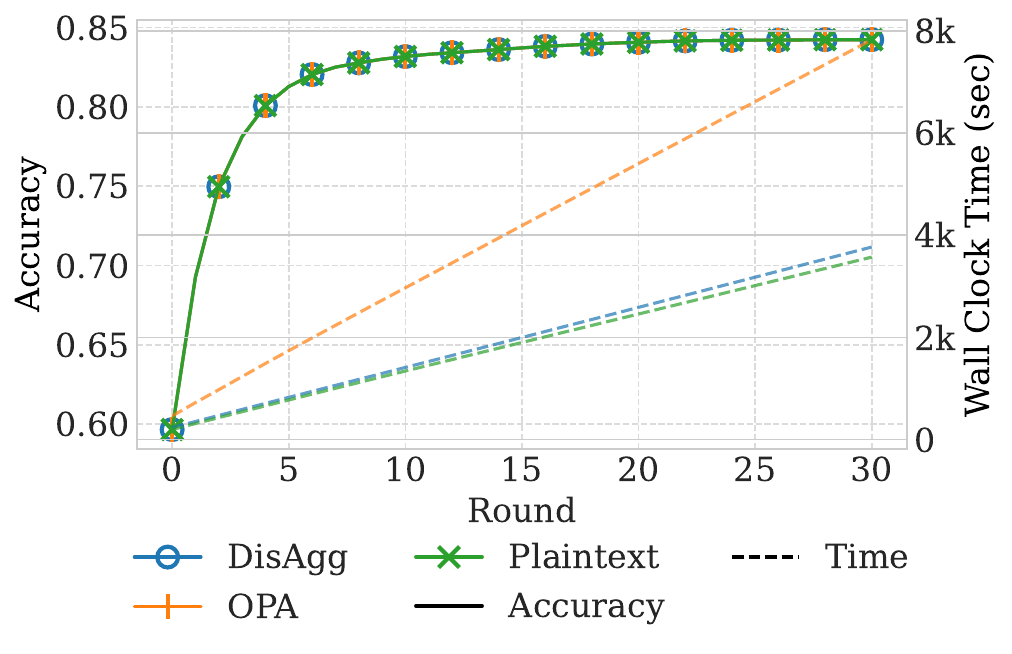}
    \vspace{-1.5em}
    \caption{CELEBA}
\end{subfigure}
\begin{subfigure}{0.48\columnwidth}
    \vspace{1.0em}
    \centering
    \includegraphics[width=\linewidth]{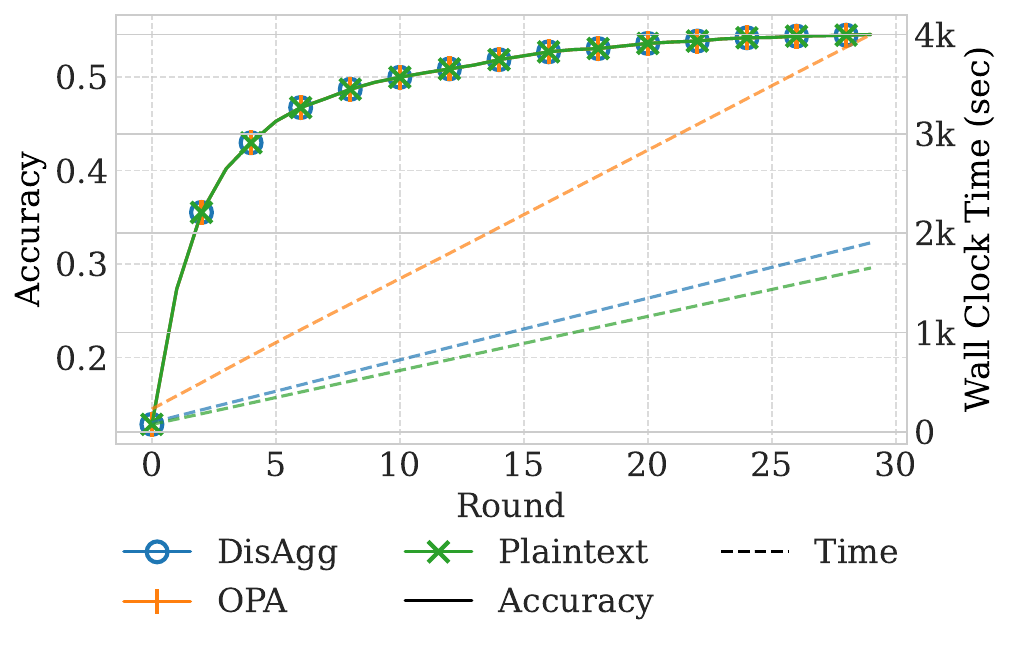}
    \vspace{-1.5em}
    \caption{CIFAR100}
\end{subfigure}
\vspace{-0.2cm}
\caption{Empirical verification for convergence analyses of \textsc{DisAgg} compared to plain-text under FL}
\label{fig:fl-training}
\vspace{-1.0em}
\end{figure}

To evaluate \textsc{DisAgg}'s suitability for plaintext recovery, we perform end-to-end multi-iteration FL using the FedAvg algorithm \cite{fedavg}, comparing simple, insecure aggregation (\textsc{Plaintext}) against secure aggregation via \textsc{DisAgg} or \textsc{OPA}. We used a diverse selection of datasets and model architectures (see Table \ref{tab:model-size}). Figure \ref{fig:fl-training} shows the validation accuracy results after 30 iterations of FL. For consistency, all settings use the same overall quantization levels. Per iteration wall clock timings are also measured and depicted on the secondary Y-axis on the right. As expected, the use of security protocols do not affect accuracy, since they have a lossless recovery of the sum. 

It is clear that \textsc{DisAgg} performs better in the cases of real-world FL scenarios, and with relatively small overhead compared to \textsc{Plaintext}. Note that training with \textsc{OPA} with $M=1.1M$ had a very high memory requirement and thus is excluded from the graphs.

\vspace{-1.0em}
\begin{table}[ht]
\centering
\caption{Number of trainable parameters for different models evaluated under plain-text recovery.}
\vspace{0.2em}
\resizebox{0.5\columnwidth}{!}{
    \begin{tabular}{ccccccccc}
    \toprule
    \textbf{Dataset} & \textbf{Model} & \textbf{Model Size} \\
    \midrule
    \textbf{MNIST} & CNN & $44.4k$ \\
    \textbf{CIFAR10} & CNN & $62k$ \\
    \textbf{CIFAR100} & TinyNet & $161k$ \\
    \textbf{SST2} & DistilBERT & $297k$ \\
    \textbf{CELEBA} & EfficientNet & $266k$ \\
    \bottomrule
    \end{tabular}
}
\vspace{-0.2em}
\label{tab:model-size}
\end{table}

\subsection{Effect of Stragglers}
\label{sec:stragglers}

In this section we present analysis on the presence of stragglers.  
A straggler is a client of the secure‑aggregation system that has an unusually long
response time. If the response time exceeds the timeout, the server may
ultimately treat a straggler as a dropout.  A dropout is any client that,
for unknown reasons, delays or even disconnects from the server.  
As explained in previous sections, \textsc{DisAgg} has a tunable tolerance to
dropouts via the $\delta$ parameter.  The protocol can handle up to a
certain fraction of dropped clients.  If a client takes a long period of time
to upload or download data, the delay may be due to a slow connection.  
Accordingly, we categorize clients into three different network speeds based
on typical mobile networks: 5G, 4G, and 3G.  Since 5G is the prevailing
option today, we evaluate 4G and 3G clients as potential stragglers.

We show that, in real‑world applications, \textsc{DisAgg} can handle these stragglers by treating 3G clients as dropouts. This is supported by the rapid decline of 3G usage, now around $10\%$ \cite{itu_mobile_nodate}, and the shutdown of 2G/3G networks in many countries \cite{amos_4g_2025}. Therefore, a relatively small percentage of these stragglers can be rejected and handled as dropouts. This allows the server to keep a lower timeout based on 4G clients while at the same time increasing the dropout tolerance. The protocol must also continue with fewer FL clients, which can affect accuracy.

To investigate these scenarios, we adjust our experimental setup to a realistic
distribution of network capabilities.  Given that 93\% of the global
population now has access to at least 4G connectivity, with 54\% having 5G
coverage \cite{amos_4g_2025}, we assume a distribution among clients’ speeds of
10\% 3G, 40\% 4G, and 50\% 5G. Since the 3G clients will be the slowest, the network will have 10$\%$ clients stragglers. 

The server must decide between two choices for stragglers: (a) wait for the stragglers to respond back and incur delay in completing the FL iteration; (b) increase $\delta$ to $0.2$, accommodating for extra dropouts and drop the stragglers.

Three distinct scenarios can then be evaluated empirically:

\begin{enumerate}[nosep]
  \item \textsc{OPA} and \textsc{DisAgg} with option (a).
  \item \textsc{OPA} with option (a) and \textsc{DisAgg} with option (b).
  \item Both \textsc{OPA} and \textsc{DisAgg} with option (b).
\end{enumerate}

\begin{figure}[ht!]
\centering
\begin{subfigure}{\columnwidth}
    \centering
    \includegraphics[width=\linewidth]{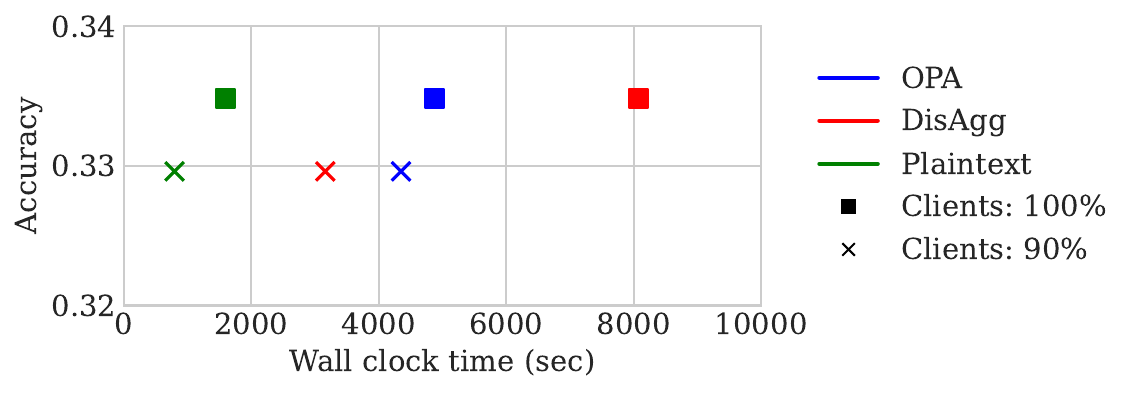}
    \vspace{-1.5em}
\end{subfigure}

\caption{Comparing \textsc{OPA} and \textsc{DisAgg} along with plaintext FL under heterogeneous clients and dropout scenarios. The accuracy depicted is after 30 iterations of training on CIFAR10.}
\label{fig:stragglers-fl-training}
\vspace{-1em}
\end{figure}

For this experiment we train a CNN vision model on the CIFAR‑10 dataset for
30 iterations with 100 clients.  Figure~\ref{fig:stragglers-fl-training}
presents the results for the three cases.  In Case 1, where both protocols
accept 3G clients, \textsc{DisAgg} is slower than \textsc{OPA} because of the
additional overhead introduced by the aggregators.  In contrast, in Cases 2
and 3, \textsc{DisAgg} outperforms \textsc{OPA}.  These three cases
reveal a trade‑off between accuracy and runtime that offers a practical
solution to the straggler problem.  By excluding the 10\% slowest clients,
\textsc{DisAgg} markedly speeds up training while incurring only a modest loss
in accuracy (about 1.5\% after 30 iterations).  In contrast, OPA does not
achieve a comparable accuracy‑runtime trade‑off: the modest speedup it
provides does not compensate for the accompanying loss in accuracy. 

\section{Conclusion}

We present \textsc{DisAgg}, a novel secure aggregation protocol for federated learning that addresses the computational and communication challenges of existing approaches through distributed aggregation. By delegating aggregation to a small subset of clients using LCC secret sharing, \textsc{DisAgg} eliminates the need for local cryptographic masking while achieving information-theoretic security. Our approach trades increased burden on Aggregators for significant reductions in computational overhead for regular clients and the server. Experiments with realistic cross-device settings of 10k+ dimensional model updates and 10k+ clients with 5G speeds demonstrate \textsc{DisAgg} achieving a minimum 3x speedup compared to the previous best protocol, \textsc{OPA}. Particularly, \textsc{DisAgg} has notable improvements in setup, client computation, and server computation, confirming that distributed aggregation can effectively balance strong security guarantees with practical efficiency for large-scale federated learning deployments.

\newpage

\bibliography{mlsys2025}
\bibliographystyle{mlsys2025}

\clearpage
\appendix

\section*{\centering\Large Appendix}
\vspace{0.5em}

\section{Related Works}
\label{app:related-works}
\textbf{FastSecAgg} \cite{fastsecagg} replaces the quadratic‑cost Shamir secret sharing of \textsc{SecAgg} with an FFT‑based multi‑secret scheme, lowering per‑client complexity to $O(N\log N)$ at the expense of reduced dropout tolerance and weaker privacy guarantees.

\textbf{LightSecAgg} \cite{lightsecagg} cuts overhead by directly reconstructing the aggregate mask of the surviving clients via Lagrange Coded Computing \cite{lagrange-coding}, preserving the privacy and dropout‑resilience of earlier schemes while still incurring multi‑phase interaction costs.  

\textbf{Flamingo} \cite{flamingo} amortizes the setup across training iterations and introduces a small decryptor committee to help the server strip aggregated masks, thereby reducing repeated communication in long‑lived deployments.  Collectively, these variants mitigate—but do not fully eliminate—the synchronization and recovery burdens that arise in large, dynamic client populations.

\textbf{Willow \cite{willow}}, a type of one-shot protocol, adopts a static, stateful committee that must execute two setup rounds followed by two decryption rounds (threshold decryption and key‑share recovery for dropped members).  It also introduces auxiliary verifier parties to audit server behavior.  As a result, Willow’s approach incurs additional communication phases and a separate verifier committee.

\textbf{TACITA \cite{tacita}} moves away from committee‑centric cryptography and incorporates a single‑server asynchronous SA protocol that performs the entire summation over encrypted client updates.  Privacy and correctness against a malicious server are enforced with threshold homomorphic encryption, zero‑knowledge proofs, and verifiable computation.  This introduces substantial overhead due to ciphertext expansion, intensive key management, and costly proof generation.  

\textbf{HierarchicalSA \cite{hierarchicalsecagg}} employs a three‑layer topology (clients $\rightarrow$ relays $\rightarrow$ server), where clients mask their updates with correlated random keys (without secret sharing) and relays aggregate their assigned subsets before forwarding partial sums.  The server then applies a linear decoding transform to recover the global sum.  Although the construction achieves information‑theoretic privacy, it does not specify dropout handling or adversarial robustness and thus is not a practically deployable SA protocol.  

\section{Theoretical Analysis}

In this section, we provide the theoretical guarantees of the \textsc{DisAgg} protocol, starting with privacy and cryptographic preliminaries followed by analyzing convergence in FL.

\subsection{Differential Privacy}
\label{app:dp}
A randomized mechanism $\mathcal{M} : \mathcal{X}^n \!\to\! \mathcal{Y}$ is $(\varepsilon,\delta)$-differentially private if for all adjacent datasets $D,D' \in \mathcal{X}^n$ differing in one record and all measurable $S \subseteq \mathcal{Y}$,
\begin{equation}
\Pr[\mathcal{M}(D)\!\in\! S] \le e^{\varepsilon}\Pr[\mathcal{M}(D')\!\in\! S] + \delta.
\end{equation}
In \emph{Central DP (CDP)}, the server aggregates raw client updates and outputs a perturbed sum $S_t + Z$, where $S_t=\sum_{i\in C_t} x_i$ is the iteration-$t$ sum over participating clients $C_t$ and $Z$ is noise calibrated to the sensitivity (e.g., Gaussian) \cite{deeplearn-dp}. In \emph{Local DP (LDP)}, each client privatizes $x_i$ with its own $(\varepsilon,\delta)$ mechanism before transmission, typically requiring more noise and reducing utility \cite{fedlearn-dp}. \emph{Distributed DP (DDP)} combines secure aggregation with distributed noise addition: each client adds an independent noise share to its update so that, when summed via SA, the aggregate attains the target CDP noise level (e.g., by splitting the Gaussian noise across clients), thereby removing the need for a trusted server to add noise centrally \cite{private-learning,dp-fedlearn-client}.

\subsection{Secret Sharing}
\label{app:ss}
We introduce threshold secret sharing because \textsc{DisAgg} splits each client’s update across $A$ Aggregators so that (i) up to $t_c$ colluding Aggregators learn nothing about any individual update, and (ii) any $t_r$ Aggregators can enable reconstruction of the needed sum. Over a prime field $\mathbb{F}_p$, we use the $(t_c,t_r,A)$ Shamir scheme \cite{shamir-secret}:
\vspace{-0.2cm}
\begin{itemize}
    \item \textsc{Share}$(s;\,t_c,t_r,A)$: sample a random polynomial $f(X)\!\in\!\mathbb{F}_p[X]$ of degree $t_r{-}1$ with $f(0)\!=\!s$, and output the shares $\{s^{(j)} := f(j)\}_{j\in[A]}$.
    \item \textsc{Coeff}$(S)$: for $S=\{i_1,\dots,i_{t_r},\dots\}\subseteq[A]$ with $|S|\!\ge\!t_r$, compute Lagrange coefficients $\lambda_{i_j} = \prod_{\zeta\in[t_r]\setminus\{j\}} \frac{i_\zeta}{i_\zeta - i_j}$ for $j=1,\dots,t_r$.
    \item \textsc{Reconstruct}$(\{s^{(j)}\}_{j\in S})$: if $|S|\!\ge\!t_r$, return $\sum_{j\in\{i_1,\dots,i_{t_r}\}} \lambda_{i_j} \cdot s^{(i_j)}$, yielding the encoded secret $s$.
\end{itemize}
\vspace{-0.2cm}
This ensures correct reconstruction and information-theoretic privacy against any coalition of at most $t_c$ malicious parties.

\subsection{Convergence Analysis}
\label{app:conv_analysis}

To analyze convergence, we adopt the standard framework used in FedAvg for non-independent and identically distributed (non-IID) data \cite{fedavg-convergence,parallel-sgd}. We outline four commonly used assumptions, followed by a fifth that models partial client participation, where $N$ out of $N_T$ total devices are selected uniformly at random without replacement in each iteration. Since our approach preserves the original FedAvg update rule, the theoretical analysis applies directly to our setting. From now on $T$ denotes the total number of local stochastic gradient updates.

\textbf{Assumption 1 (Smoothness):} Each local objective $F_k$ is $L$-smooth, i.e., for all vectors $v, w \in \mathbb{R}^d$, and for all $k \in \{1, \dots, N_T\}$:
\begin{equation}
F_k(v) \leq F_k(w) + \langle \nabla F_k(w), v - w \rangle + \frac{L}{2} \|v - w\|_2^2.
\end{equation}

\textbf{Assumption 2 (Strong Convexity):} Each $F_k$ is $\mu$-strongly convex, meaning that for all $v, w \in \mathbb{R}^d$:
\begin{equation}
F_k(v) \geq F_k(w) + \langle \nabla F_k(w), v - w \rangle + \frac{\mu}{2} \|v - w\|_2^2.
\end{equation}

\textbf{Assumption 3 (Bounded Gradient Variance):} Let $\xi_t^k$ be a stochastic sample drawn uniformly at random from the local dataset of client $k$. Then the variance of the stochastic gradients is bounded:
\begin{equation}
\mathbb{E} \left[ \left\| \nabla F_k(w_t^k, \xi_t^k) - \nabla F_k(w_t^k) \right\|^2 \right] \leq \sigma_k^2.
\end{equation}

\textbf{Assumption 4 (Bounded Gradient Norm):} The expected squared norm of the stochastic gradients is uniformly bounded:
\begin{equation}
\mathbb{E} \left[ \left\| \nabla F_k(w_t^k, \xi_t^k) \right\|^2 \right] \leq G^2,
\end{equation}
for all $k = 1, \dots, N_T$ and $t = 1, \dots, T-1$.

\textbf{Assumption 5 (Partial Client Participation):} At each iteration $t$, a subset $S_t \subseteq [N_T]$ of $N$ clients is selected uniformly at random without replacement. The data distribution is assumed to be balanced, i.e., $p_k = \frac{1}{N_T}$ for all $k$. The server aggregates client models using:
\begin{equation}
w_t \leftarrow \frac{N_T}{N} \sum_{k \in S_t} p_k w_t^k.
\end{equation}

\textbf{Quantifying Non-IIDness:}
To capture the degree of data heterogeneity across clients, let $F^*$ denote the global minimum of $F$, and $F_k^*$ the minimum of each local objective $F_k$. We define the heterogeneity gap as:
\begin{equation}
\Gamma = F^* - \sum_{k=1}^{N_T} p_k F_k^*
\end{equation}
When data are IID, $\Gamma \to 0$ as the number of samples increases. In contrast, a nonzero $\Gamma$ indicates Non-IIDness, with its magnitude reflecting the extent of distributional divergence across clients.

Under these assumptions, convergence guarantee can be established for the FedAvg algorithm as follows.

\textbf{Theorem 2:}
Let Assumptions 1 to 4 hold, with constants $L$, $\mu$, $\sigma_k$, and $G$ defined therein. Define the condition number $\kappa = \frac{L}{\mu}$, $\xi = \max\{8\kappa, E\}$, and set the learning rate to $\eta_t = \frac{2}{\mu(\xi + t)}$. Assume Assumption 5 holds, and let $C = \frac{N_T-N}{N_T-1} \cdot \frac{4E^2G^2}{N}$. Then:
\vspace{-0.2cm}
\begin{equation}
\begin{aligned}
    \mathbb{E}[F(w_T)] - F^* \leq \\\frac{\kappa}{\xi + T - 1} &\left( \frac{2(B + C)}{\mu} + \frac{\mu \xi}{2} \mathbb{E}\|w_1 - w^*\|^2 \right)
\end{aligned}
\end{equation}

where
\vspace{-0.2cm}
\begin{equation}
B = \sum_{k=1}^{N_T} p_k^2 \sigma_k^2 + 6L\Gamma + 8(E - 1)^2 G^2.
\end{equation}

\noindent
Here, $E$ is the number of local iterations per one training iteration.

\begin{table*}[ht]
\centering
\caption{Extended computational and communication complexity for \textsc{OPA} and \textsc{DisAgg}. $N$ is the number of selected clients, $A$ is the Aggregator group size, $M$ is the model size, $\rho$ is the packing factor of \textsc{OPA}, and $\lambda$ is the key size.}
\footnotesize
\renewcommand{\arraystretch}{1.2}
\setlength{\tabcolsep}{6pt}
\begin{tabular}{|c|c|c|c|}
\hline
& \textbf{Client} & \textbf{Committee} & \textbf{Server} \\
\hline
\textbf{OPA (Comp.)} & $O(\lambda M + \tfrac{\lambda}{\rho} A \log A + A + A^2)$ & $O(\lambda N)$ & $O(\lambda M + NM + \tfrac{\lambda}{\rho} A \log A + A^2)$ \\
\hline
\textbf{OPA (Comm.)} & $O(M + \tfrac{\lambda}{\rho}A + A)$ & $O(N + N \tfrac{\lambda}{\rho} + \tfrac{\lambda}{\rho})$ & $O(NA + NM + 2AN \tfrac{\lambda}{\rho} + A \tfrac{\lambda}{\rho})$ \\
\hline
\textbf{DisAgg (Comp.)} & $O(\tfrac{M}{\rho}AlogA + A^2)$ & $O(N\tfrac{M}{\rho})$ & $O(\tfrac{M}{\rho}A \log A + A^2)$ \\
\hline
\textbf{DisAgg (Comm.)} & $O(\tfrac{M}{\rho}A + A)$ & $O(\tfrac{NM}{\rho} + N + \tfrac{M}{\rho})$ & $O(2AN\tfrac{M}{\rho} + NA + \tfrac{M}{\rho}A)$ \\
\hline
\end{tabular}
\label{tab:opa_disagg_extended}
\vspace{-0.3cm}
\end{table*}

\begin{table*}[ht]
\centering
\caption{Altered complexity table for \textsc{OPA} and \textsc{DisAgg}, using $O(A^2)$ instead of $O(A \log A)$ for the secret sharing and reconstruction operations. This setting corresponds to the implementation complexity used in our simulations.}
\footnotesize
\renewcommand{\arraystretch}{1.2}
\setlength{\tabcolsep}{6pt}
\begin{tabular}{|c|c|c|c|}
\hline
& \textbf{Client} & \textbf{Committee} & \textbf{Server} \\
\hline
\textbf{\textsc{OPA} (Comp.)} & $O(\lambda M + \tfrac{\lambda}{\rho}A^2 + A + A^2)$ & $O(\lambda N)$ & $O(\lambda M + NM + \tfrac{\lambda}{\rho}A^2 + A^2)$ \\
\hline
\textbf{\textsc{OPA} (Comm.)} & $O(M + \tfrac{\lambda}{\rho}A + A)$ & $O(N + N\tfrac{\lambda}{\rho} + \tfrac{\lambda}{\rho})$ & $O(NA + NM + 2AN\tfrac{\lambda}{\rho} + A\tfrac{\lambda}{\rho})$ \\
\hline
\textbf{DisAgg (Comp.)} & $O(\tfrac{M}{\rho}A^2 + A^2)$ & $O(N\tfrac{M}{\rho})$ & $O(\tfrac{M}{\rho}A^2 + A^2)$ \\
\hline
\textbf{DisAgg (Comm.)} & $O(\tfrac{M}{\rho}A + A)$ & $O(N\tfrac{M}{\rho} + N + \tfrac{M}{\rho})$ & $O(2AN\tfrac{M}{\rho} + NA + \tfrac{M}{\rho}A)$ \\
\hline
\end{tabular}
\label{tab:opa_disagg_altered}
\end{table*}

\textbf{Remark 1 (General Convex and Non-Convex Objectives):}
While Theorem~1 addresses the strongly convex case, analogous convergence results can be established for general convex and non-convex objectives by leveraging the analysis framework from previous work.

\textbf{Remark 2 (Client-Specific Dropout Behavior):}
In realistic scenarios, clients may have varying dropout probabilities, which violates the assumption of uniformly random client selection and complicates the theoretical guarantees. Nonetheless, empirical evidence suggests that FedAvg continues to converge even under such heterogeneous participation patterns.

\textbf{Remark 3 (Choice of $E$):}
As shown in \cite{fedavg-convergence}, the total communication rounds $T / E$ is a non-monotonic function of $E$, initially decreasing and then increasing. This suggests that overly small or large values of $E$ may incur high communication costs, and that an optimal choice of $E$ exists.

\textbf{Remark 4 (Choice of $N$):}
The convergence rate exhibits only weak dependence on the number of active clients $N$. This implies that the participation ratio $N/N_T$ can be kept small to mitigate straggler effects, without significantly impacting convergence.

\textbf{Remark 5 (Error Rate):}
The convergence rate is $\mathcal{O}(1/T)$, meaning that the expected gap to the optimum satisfies:
\begin{equation}
\mathbb{E}[F(w_T)] - F^* \leq \frac{\text{const.}}{T}
\end{equation}
Hence, increasing the total number of steps $T$ leads to a linear decrease in worst-case error.

Next we point out the analysis of \cite{fedavg-convergence} that diminishing learning rates are crucial for the convergence of FedAvg in the Non-IID setting.

\textbf{Theorem 3 (Effect of Learning Rate Decay):}
FedAvg does not necessarily converge to the optimal solution under fixed learning rates when $E > 1$. Let $\tilde{w}^*$ denote the solution obtained by FedAvg with constant step size $\eta$, and let $w^*$ be the optimal point. Then,
\begin{equation}
    \|\tilde{w}^* - w^*\|_2 = \Omega((E - 1)\eta) \cdot \|w^*\|_2
\end{equation}
up to constant factors. This construction highlights the necessity of diminishing step sizes in the Non-IID setting.

\textbf{Remark 6 (Learning Rate Schedule):}
When using a decaying learning rate and $E > 1$, FedAvg converges to the optimum. In contrast, with a fixed learning rate and $E > 1$, the algorithm does not converge to the optimum.

\section{Complexity Analysis}
\label{app:complexity_breakdown}
\subsection{Breakdown of Theoretical Complexities}
\textbf{Regular client:} A client has a pair of private-public keys for communication with all Aggregators, and exchanges public keys with the Aggregators via the server. It sends its own with $\mathcal{O}(1)$ time complexity and receives the public key of all Aggregators with $\mathcal{O}(A)$ complexity, where $A < N$ is the size of the Aggregator group. Each client generates the shares of their model with complexity $\mathcal{O}(M \log A)$, where $M$ is the model size (i.e., the number of model parameters). This complexity arises from the use of Fast Fourier Transform (FFT) \cite{fft-algorithm} which can be used in theory for the polynomial evaluation. The total communication cost is $\mathcal{O}(M + A)$.

\textbf{Aggregator:} An Aggregator, similarly to the other clients, participates in the one-time key exchange: it broadcasts a single public key and (via the server) receives $N$ client public keys to set up encrypted channels, yielding the $N$ term. It then receives one secret share of size $\approx M/A$ from each of the $N$ clients, so the total payload per Aggregator is $N \cdot (M/A)$ field elements, giving communication $\mathcal{O}((M/A + N)$. To compute its partial sum, the Aggregator adds these $N$ shares entrywise (each of length $M/A$), which costs $\mathcal{O}(N \cdot M/A)$ arithmetic.

\textbf{Server:} The server transfers all public keys between $N$ clients and $A$ Aggregators with cost $\mathcal{O}(NA)$, the $N$ models each of size $M$ with $\mathcal{O}(NM)$, and the aggregated parts from the Aggregators with $\mathcal{O}(M)$. The total communication cost is $\mathcal{O}(NM + NA)$. For the reconstruction of the aggregated model, the computation complexity is the same as the client's sharing cost, in order to combine the shares: $\mathcal{O}(M \log A)$.

\subsection{Refining Complexities}
\label{app:complexity_extended}

In this section, we provide the detailed comparison table of the extended complexities for \textsc{DisAgg} and \textsc{OPA}, as discussed in main paper (Table~\ref{tab:opa_disagg_extended}). We also provide in Table~\ref{tab:opa_disagg_altered} a slightly modified version, where the complexity of secret sharing and reconstruction for both schemes is set to $O(A^2)$ instead of $O(A \log A)$. This alteration reflects the complexity used in our simulations for implementation simplicity. Since this change affects both protocols symmetrically, it does not influence the conclusions of the comparison.

\subsection{Alternative Network Speeds}
\label{app:alt-network}

We complement the main 5G-oriented evaluation with experiments under reduced client bandwidth representative of 4G and 3G connectivity. Since \textsc{DisAgg} incurs additional Aggregator-facing communication, lower uplink capacity disproportionately affects its relative performance. Nevertheless, \textsc{DisAgg} continues to achieve a (reduced) speedup over \textsc{OPA} in both alternative settings.

Figures~\ref{fig:optimal-a-heatmap-4g} and \ref{fig:optimal-a-heatmap-3g} report (for each network profile) (i) the optimal number of Aggregators $A$ selected independently for \textsc{OPA} and \textsc{DisAgg} across $(M,N)$ pairs under the same optimization criterion as in the main text, and (ii) the corresponding relative speedup of \textsc{DisAgg} over \textsc{OPA}. The 4G configuration uses $200\,\text{kBps}$ upload and $2\,\text{MBps}$ download per client; the 3G configuration uses $50\,\text{kBps}$ upload and $500\,\text{kBps}$ download per client. The server bandwidth is held fixed at the 5G setting. As expected, reduced client bandwidth narrows the performance gap, yet \textsc{DisAgg} still exhibits a positive speedup in these regimes.

\subsection{Aggregator Stragglers}
\label{app:stragglers}

\begin{figure*}[!htb]
    \centering
    \includegraphics[width=\linewidth]{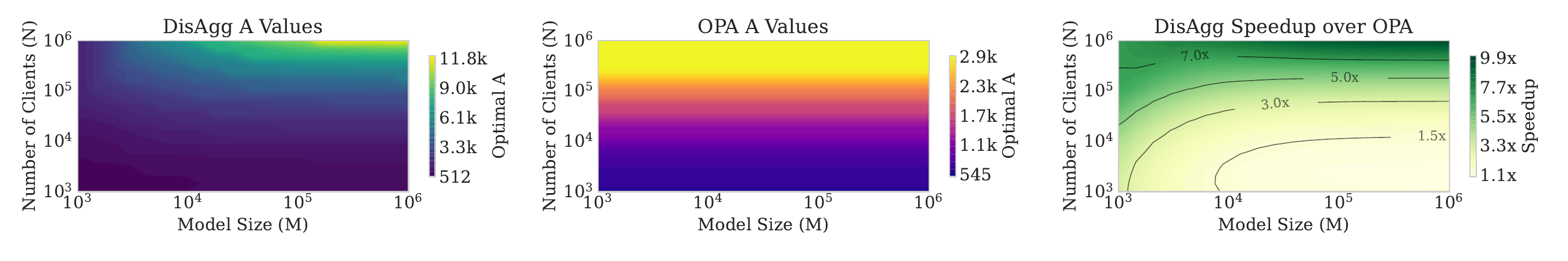}
    \vspace{-1.2cm}
    \caption{Contour plot showing the optimal number of Aggregators $A$ for \textsc{DisAgg} and \textsc{OPA} and the resulting speedup of \textsc{DisAgg} over \textsc{OPA} across $(M, N)$ under a 4G client connectivity assumption (200 kBps upload / 2 MBps download), with $k=0.3$ and $k_{\text{comp}}=0.66$.}
    \label{fig:optimal-a-heatmap-4g}
\end{figure*}

\begin{figure*}[!htb]
    \centering
    \includegraphics[width=\linewidth]{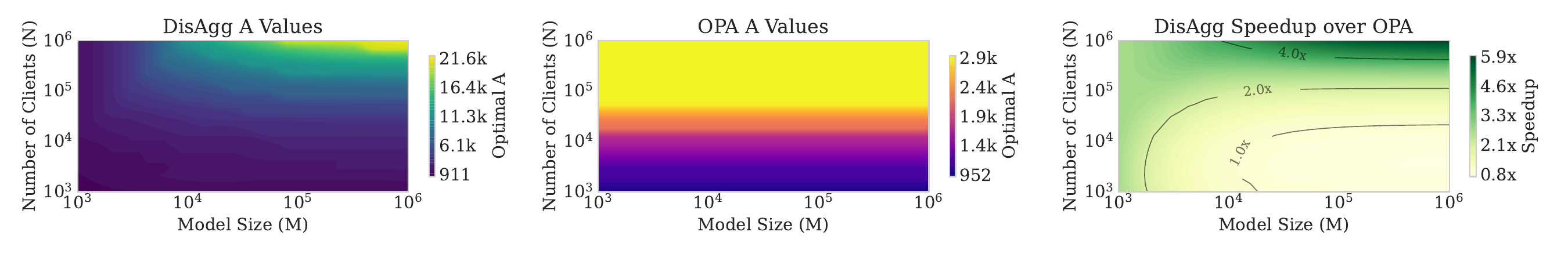}
    \vspace{-1.2cm}
    \caption{Contour plot showing the optimal number of Aggregators $A$ for \textsc{DisAgg} and \textsc{OPA} and the resulting speedup of \textsc{DisAgg} over \textsc{OPA} across $(M, N)$ under a 3G client connectivity assumption (50 kBps upload / 500 kBps download), with $k=0.3$ and $k_{\text{comp}}=0.66$.}
    \label{fig:optimal-a-heatmap-3g}
\end{figure*}

Assuming that server bandwidth is not a bottleneck (as multiple servers can be deployed in parallel), the per-round latency is determined by the slowest surviving Aggregator. This effect can be equivalently modeled by treating Aggregators as clients with heterogeneous network speeds (e.g., 3G/4G/5G). We evaluate the expected speedups under such heterogeneous settings in Figures \ref{fig:optimal-a-heatmap-4g} and \ref{fig:optimal-a-heatmap-3g} as discussed above, and further extend our experimental evaluation on heterogeneous clients in Figure \ref{fig:hetero-comparison-main}. Even when 40\% of Aggregators use 4G connections instead of 5G, \textsc{DisAgg} continues to achieve a speedup over \textsc{OPA}. Moreover, due to its high dropout tolerance (see Section \ref{sec:dropout-tolerance}), \textsc{DisAgg} can accommodate a finite round timeout to handle extreme stragglers, and is therefore not significantly impacted by Aggregator stragglers or client heterogeneity.

\subsection{Minimum Aggregators Analysis}
\label{app:minimum_a}

\begin{figure*}[ht]
    \centering
    \includegraphics[width=\linewidth]{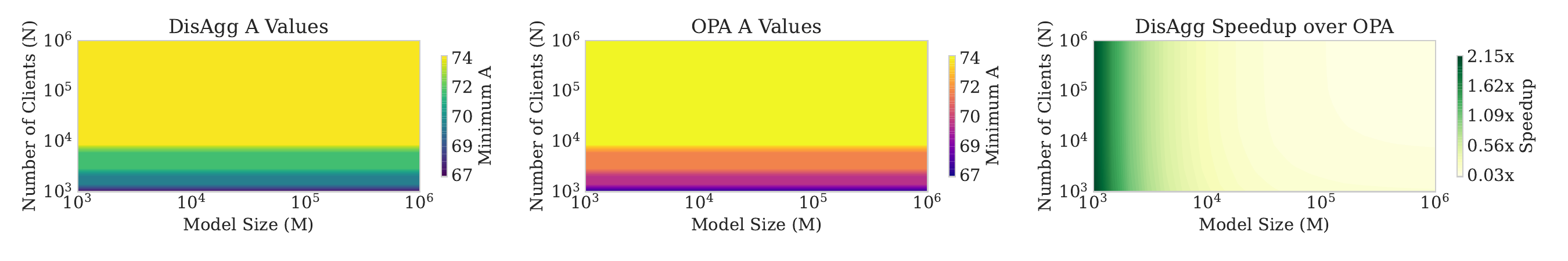}
    \vspace{-3.0em}
    \caption{Contour plot showing the minimum number of Aggregators $A$ (calculated for minimum packing factor $\rho=1$) for \textsc{DisAgg} and \textsc{OPA} and the resulting speedup of \textsc{DisAgg} over \textsc{OPA} across $(M, N)$ under a 5G client connectivity assumption (2 MBps upload / 20 MBps download), with $k=0.3$ and $k_{\text{comp}}=0.66$.}
    \label{fig:optimal-a-heatmap-minA}
\end{figure*}

\begin{figure*}[ht]
    \centering
    \includegraphics[width=\linewidth]{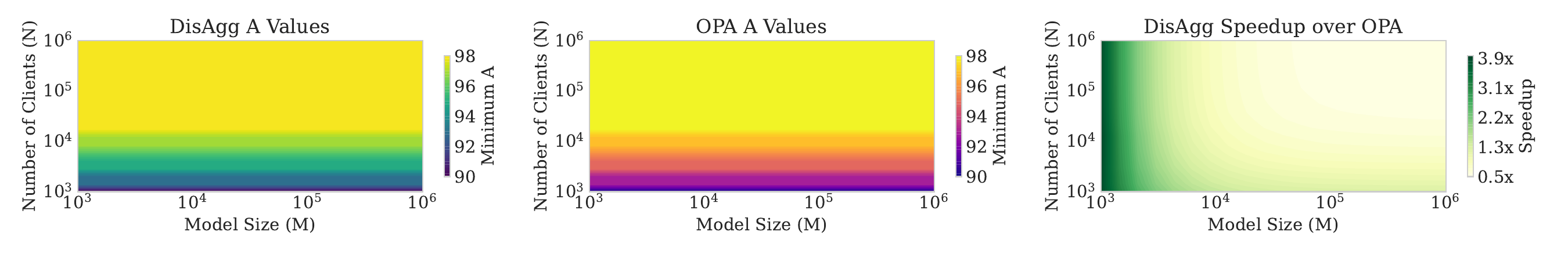}
    \vspace{-3.0em}
    \caption{
    Contour plot showing the minimum number of Aggregators $A$ for \textsc{DisAgg} and \textsc{OPA} and the resulting speedup of \textsc{DisAgg} over \textsc{OPA} across $(M, N)$ under a 5G client connectivity assumption (2~MBps upload / 20~MBps download), with $k=0.3$ and $k_{\text{comp}}=0.66$. The number of Aggregators are computed for packing factor $\rho = 16$, matching \textsc{OPA}'s configuration.}
    \label{fig:optimal-a-heatmap-rho16}
\end{figure*}

In this section, we present contour plots that illustrate the minimum number of Aggregators $A$ (calculated for minimum packing factor $\rho=1$) across $(M, N)$ pairs for both \textsc{DisAgg} and \textsc{OPA} under the 5G setting. The improvement of \textsc{DisAgg} over \textsc{OPA} is shown on the right of Figure~\ref{fig:optimal-a-heatmap-minA}, where performance gains range from $0.03\times$ to $2.15\times$. To align with the configuration used in the \textsc{OPA} paper~\cite{opa}, we additionally report results for $\rho=16$, corresponding to \textsc{OPA}’s packing choice. Figure~\ref{fig:optimal-a-heatmap-rho16} presents this comparison under identical packing assumptions, with speedup factor from $0.5\times$ to $3.9\times$.

\vspace{-0.5em}
\subsection{Packing Optimization}
A potential optimization, left as future work, is to pack multiple 32-bit secrets into a single 128-bit field element. Since summing $N$ client inputs requires an additional $\lceil \log_2 N \rceil$ bits of headroom to prevent overflow, one could encode two 32-bit model parameters (plus the required padding) within each 128-bit field element, effectively halving the Aggregator download burden, while maintaining correctness and security.

\begin{figure*}[!htb]
    \centering
    \includegraphics[width=\linewidth]{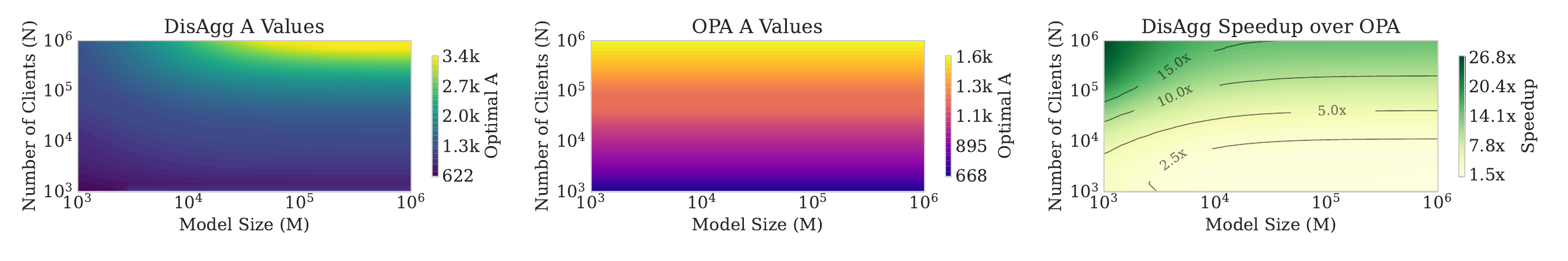}
    \vspace{-1.2cm}
    \caption{Contour plot showing the optimal number of Aggregators $A$ for \textsc{DisAgg} and \textsc{OPA} and the speedup of \textsc{DisAgg} over \textsc{OPA} across $(M, N)$ under the BFT constraint ($\gamma + \delta \le 1/3$) and a 5G client connectivity assumption, with $\gamma = \delta = 0.1$ and $k_{\text{comp}}=0.66$.}
    \label{fig:optimal-a-heatmap-bft}
\end{figure*}

\vspace{-1.0em}
\subsection{Byzantine Fault Tolerance}
\label{app:bft}

We present theoretical complexity plots under the Byzantine fault-tolerance (BFT) constraint, where the minimum committee size is chosen to satisfy $\gamma + \delta \le 1/3$ (as discussed in main paper). For illustration, we instantiate $\gamma = \delta = 0.1$. Figure~\ref{fig:optimal-a-heatmap-bft} reports the optimal number of Aggregators $A$ for \textsc{DisAgg} and \textsc{OPA} across $(M,N)$, together with the speedup of \textsc{DisAgg} over \textsc{OPA}, under the BFT constraint. As expected, enforcing BFT increases the required committee size and leads to an overall reduction in the relative performance of \textsc{DisAgg} compared to the non-BFT case.

\vspace{-0.5em}
\section{Additional Experiments}
\subsection{Training and Quantization}

\begin{figure*}[!htb]
\centering
\begin{subfigure}{0.19\linewidth}
    \centering
    \includegraphics[width=0.975\linewidth]{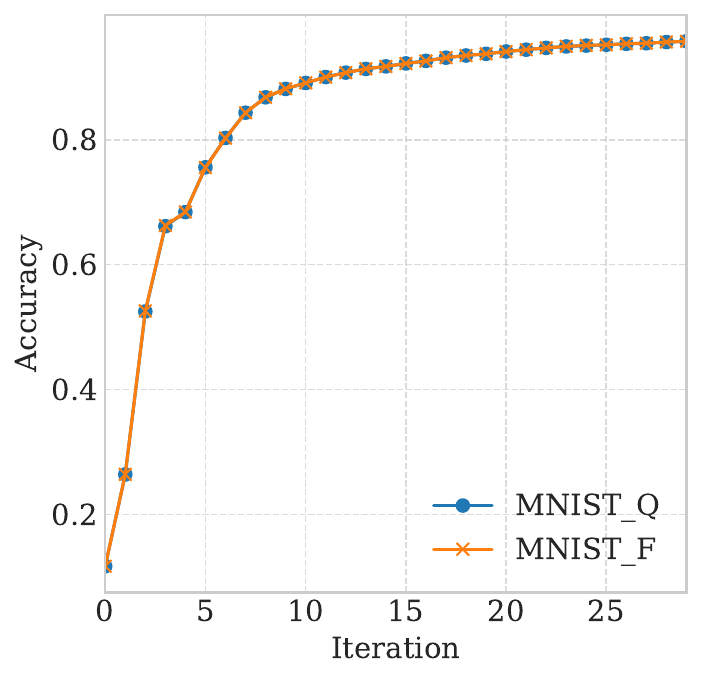}
    \caption{MNIST}
\end{subfigure}
\begin{subfigure}{0.19\linewidth}
    \centering
    \includegraphics[width=\linewidth]{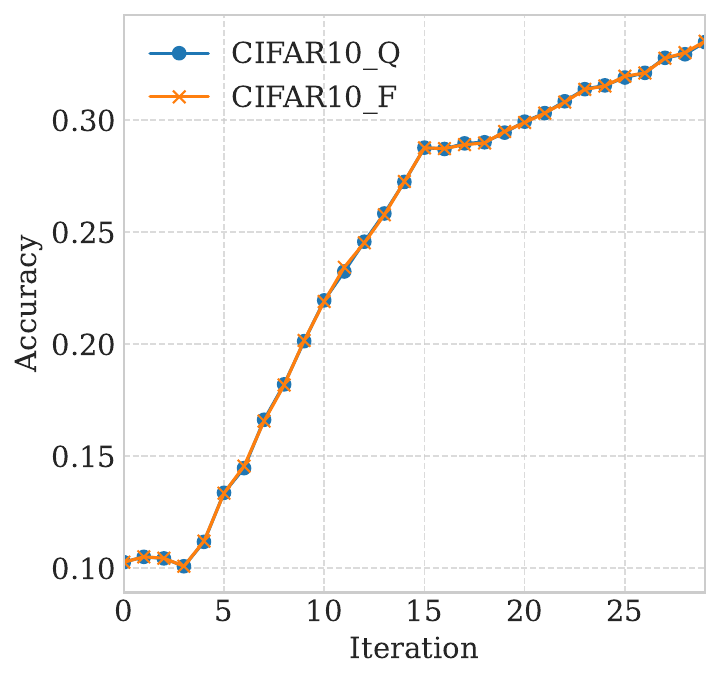}
    \caption{CIFAR10}
\end{subfigure}
\begin{subfigure}{0.19\linewidth}
    \centering
    \includegraphics[width=\linewidth]{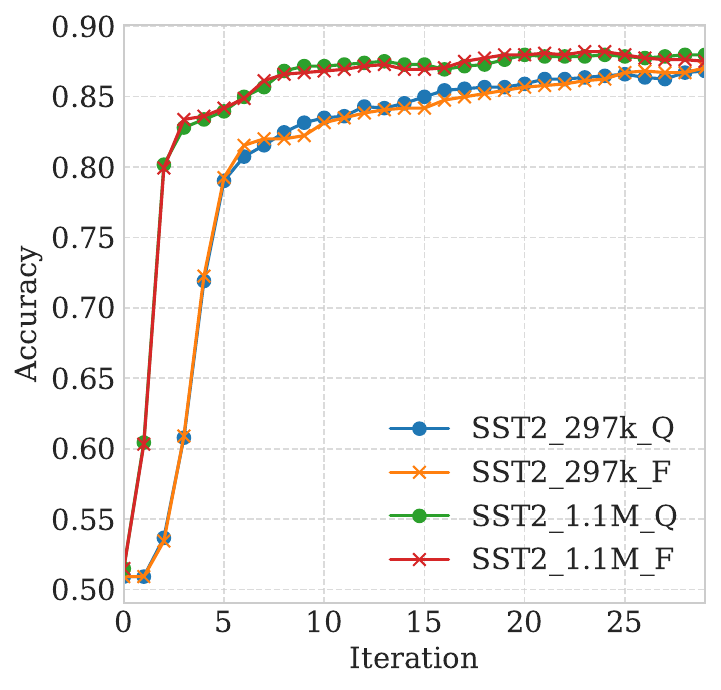}
    \caption{SST2}
\end{subfigure}
\begin{subfigure}{0.19\linewidth}
    \centering
    \includegraphics[width=\linewidth]{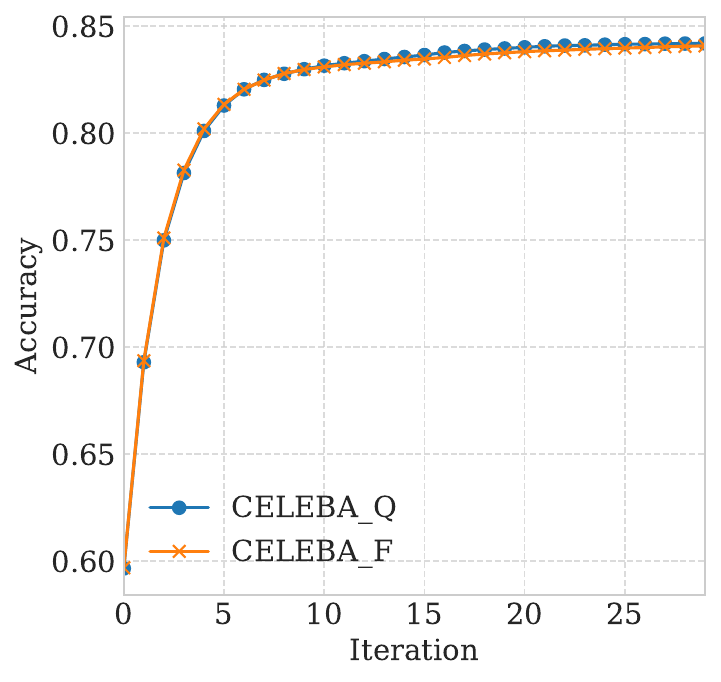}
    \caption{CELEBA}
\end{subfigure}
\begin{subfigure}{0.19\linewidth}
    \centering
    \includegraphics[width=0.975\linewidth]{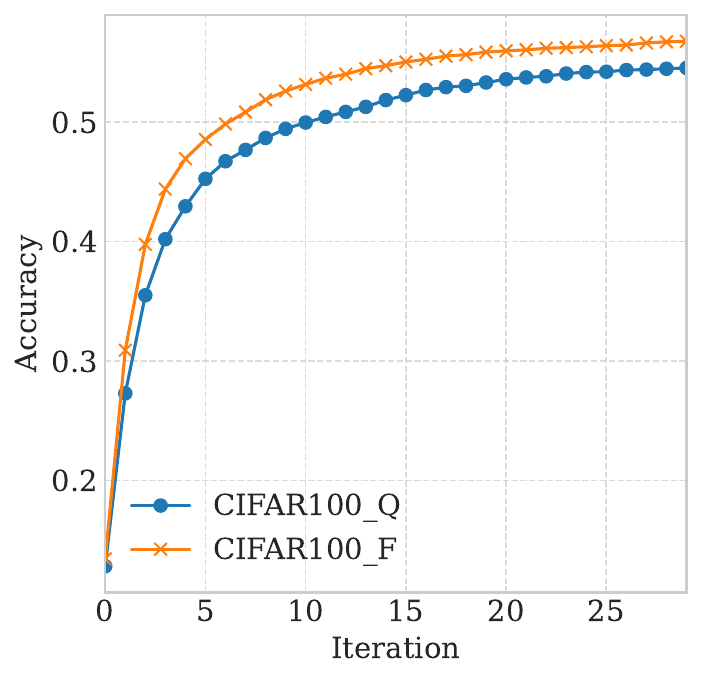}
    \caption{CIFAR100}
\end{subfigure}
\vspace{-0.25em}
\caption{Training accuracy with different datasets and model using plaintext FL. Q denotes the use of quantization required for cryptographic primitives in secure aggregation. F denotes floating point precision as used in standard FL.}
\label{fig:train-fl-quant}
\end{figure*}

We conducted experiments to test the effect of quantization in our secure aggregation protocol. Figure \ref{fig:train-fl-quant} shows the training results for the models and datasets that are used in the section 5.4. In this setup we train using only plaintext FL but with two options. The first is to keep the quantization as in the secure aggregation protocols, and the second is to do the training directly with the model parameters as floats, as the nornal FL does, without security. The first is denoted with the letter 'Q' for quantization depicted on the dataset names and the latter with 'F' for floats. The plaintext field we used for quantization was 53 bits and as we can see in this Figure \ref{fig:train-fl-quant} the effect in the final accuracy for every round is negligible, with an exception for CIFAR100 dataset with TinyNet model. In this case the difference in the final accuracy is $\approx4\%$. The main cause for this deviation is the clipping of the model parameters prior to translating floats to integers. The clipping range is set to [-2.0, 2.0] for all models.

\vspace{-0.5em}
\subsection{Training Timings}

We measured the timings for every phase during one round of FL for the secure aggregation protocols \textsc{DisAgg} and \textsc{OPA} along with simple aggregation \textsc{Plaintext}. The results are presented in Tables \ref{tab:train-nlp-fl} and \ref{tab:train-efficientnet-fl}, reporting the timings for training a DistilBERT model with the SST2 dataset, and EfficientNet model with the CelebA dataset respectively. \textsc{DisAgg} only adds up to 20$\%$ overhead while \textsc{OPA} can add up to 190$\%$ overhead compared to \textsc{Plaintext}.

\vspace{-0.5em}
\subsection{Grid Search for Large-scale Experiments}
We present a grid search over the parameter $\rho$ for both \textsc{OPA} and \textsc{DisAgg} to produce the results reported in Table \ref{tab:opa-vs-disagg-100k}. The grid search highlights the importance of tuning the $\rho$ for each method. In contrast, $\rho$ was kept fixed in \cite{opa}.

\begin{table*}[!htb]
\caption{Per stage timings in seconds for \textsc{DisAgg}, \textsc{OPA} and \textsc{Plaintext} (simple aggregation) during one FL iteration. A DistilBERT model with 297k trainable parameters is trained on the SST2 dataset. \textsc{DisAgg} has a marginal overhead of 3.1$\%$ over \textsc{Plaintext}.}
\centering
\resizebox{0.95\textwidth}{!}{
    \begin{tabular}{ccccccccc}
    \toprule
    \textbf{Method} & \textbf{Setup} & \textbf{Client Comm} & \textbf{Committee Comm} & \textbf{Client Comp} & \textbf{Committee Comp} & \textbf{Server Comp} & \textbf{Total w/out Setup} & \textbf{Total} \\
    \midrule
    \textsc{OPA} & 123.59& 1.43& 0.02& 163.76& 1e-3& 72.96& 238.17& 361.76
\\
    \textsc{DisAgg} & 0.75& 7.07& 1.79& 66.27& 0.09& 0.19& 75.41& 76.16
\\
    \textsc{Plaintext} & -& 0.64& -& 73.13& -& 0.11& 73.89& 73.89
\\
    \bottomrule
    \end{tabular}
}
\vspace{-0.2em}
\label{tab:train-nlp-fl}
\end{table*}

\begin{table*}[!htb]
\caption{Per stage timings in seconds for \textsc{DisAgg}, \textsc{OPA} and \textsc{Plaintext} (simple aggregation) during one FL iteration. An EfficientNet model with 266k parameters is trained on the CelebA dataset. \textsc{DisAgg} has a marginal overhead of 20$\%$ over \textsc{Plaintext}.
}
\centering
\resizebox{0.95\textwidth}{!}{
    \begin{tabular}{ccccccccc}
    \toprule
    \textbf{Method} & \textbf{Setup} & \textbf{Client Comm} & \textbf{Committee Comm} & \textbf{Client Comp} & \textbf{Committee Comp} & \textbf{Server Comp} & \textbf{Total w/out Setup} & \textbf{Total} \\
    \midrule
    \textsc{OPA} & 109.81& 1.29& 0.02& 291.68& 2e-3& 65.58& 358.57& 468.38
\\
    \textsc{DisAgg} & 0.76& 6.30& 1.61& 184.38& 0.08& 0.17& 192.54& 193.30
\\
    \textsc{Plaintext} & -& 0.57& -& 160.34& -& 0.12& 161.03& 161.03
\\
    \bottomrule
    \end{tabular}
}
\vspace{0.3em}
\label{tab:train-efficientnet-fl}
\end{table*}

\begin{figure}[!h]
\centering
\begin{subfigure}{\columnwidth}
    \flushleft
    \vspace{-0.2cm}
    \includegraphics[width=0.70\columnwidth]{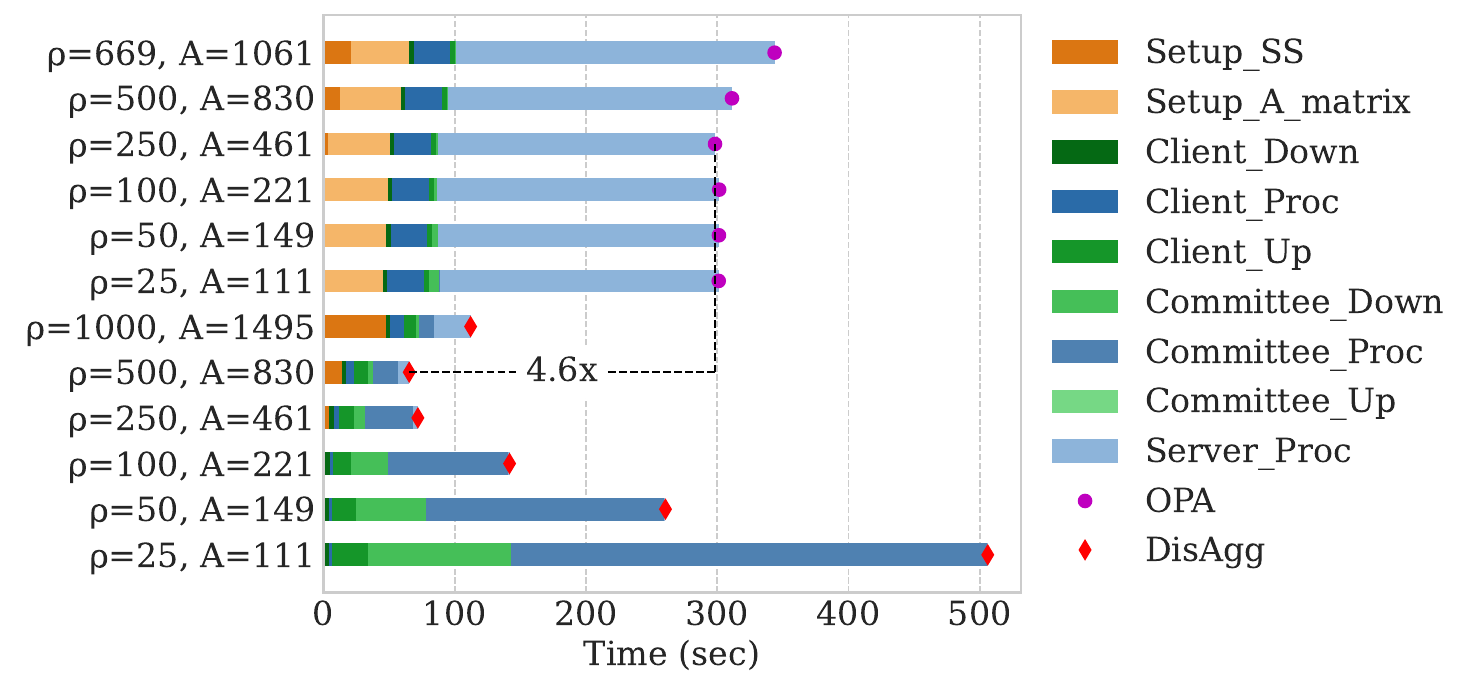}
    \vspace{0.2cm}
\end{subfigure}
\begin{subfigure}{\columnwidth}
    \flushleft
    \includegraphics[width=0.49\columnwidth]{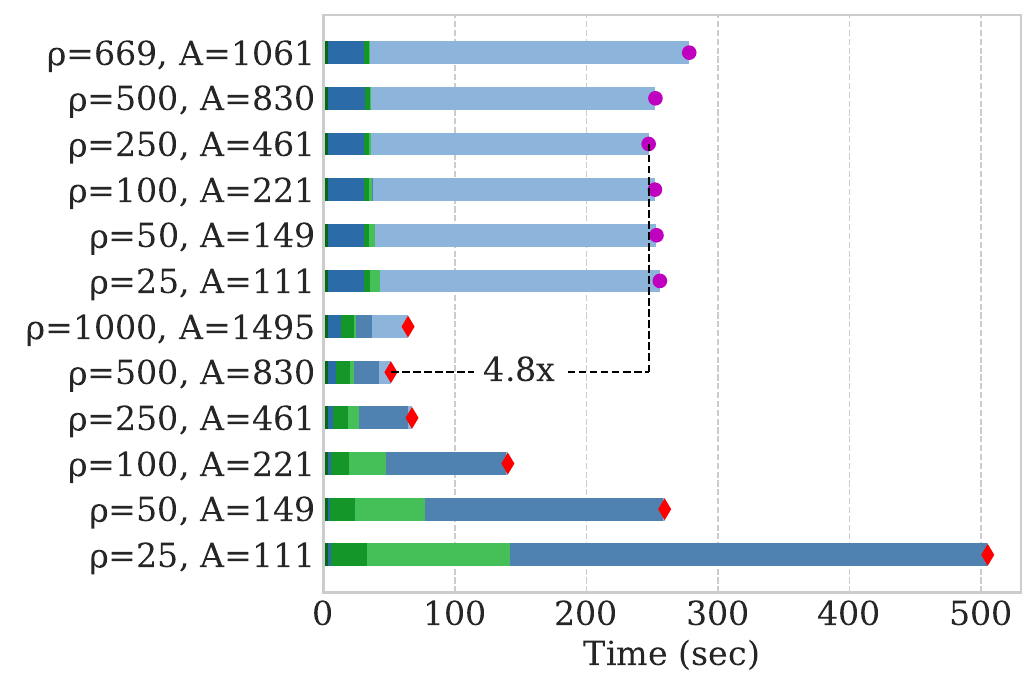}
    \vspace{-1.0em}
\end{subfigure}
    \vspace{-0.5em}
    \caption{Speedup of \textsc{DisAgg} over \textsc{OPA} for one FL iteration with $M=N=100k$ using a grid search of $\rho$ as described in Section \ref{sec:disagg-vs-opa}. Equations \ref{eq:cdf_corrupted}--\ref{eq:thresholds} are used to calculate $A$ for a given $\rho$.}
    \label{fig:100k-rho-analysis}
\vspace{-1.0em}
\end{figure}

\begin{figure}[!ht]
    \centering
    \includegraphics[width=0.94\columnwidth]{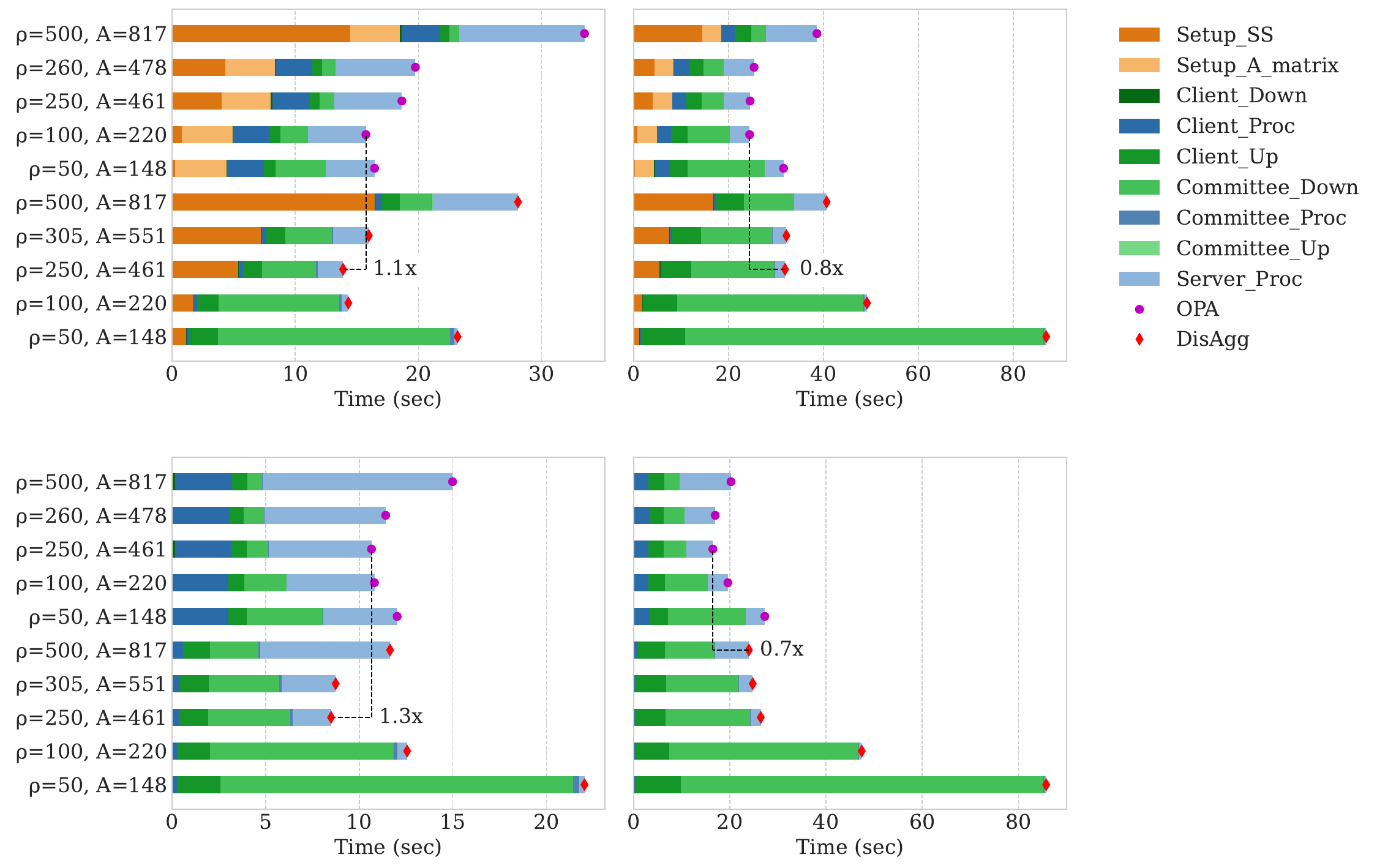}
    \vspace{-0.25em}
    \caption{Time comparison for one FL iteration of \textsc{DisAgg} and \textsc{OPA} with ($60\%$ 5G, $40\%$ 4G) clients on the left, ($60\%$ 5G, $40\%$ 3G) on the right and a shared y-axis for each row. The top row includes the setup times while the bottom row excludes it. DisAgg is within $\pm30\%$ of OPA timings under such conditions.}
    \label{fig:hetero-comparison-main}
    \vspace{-1.0em}
\end{figure}

\subsection{Effect of Heterogeneity}
\label{app:hetero-stragglers}
Section~\ref{sec:stragglers} discusses the effects of stragglers on  \textsc{DisAgg}'s performance. In this section, we compare the effect of network heterogeneity on both \textsc{DisAgg} and \textsc{OPA}, by analyzing the effects of clients having a distribution of network speeds. Figure \ref{fig:hetero-comparison-main} presents cases with (60$\%$ 5G, 40$\%$ 4G) and (60$\%$ 5G, 40$\%$ 3G) client speed distributions. In the first case, \textsc{DisAgg} is still faster than \textsc{OPA} by 10-30$\%$, whereas in the second case of having 40$\%$ clients with 3G connectivity, delays on the Aggregators' download of secret shares in \textsc{DisAgg} lead to a 20-30$\%$ slowdown compared to \textsc{OPA}. Given that $93\%$ of the global population has access to at least 4G connectivity today \cite{amos_4g_2025}, the second scenario represents an extremely unlikely case. In practical scenarios, the chances of encountering 3G clients may be no more than 10$\%$; the server could either never select such clients initially or regard them as dropouts later. As shown in Section~\ref{sec:stragglers}, in such cases, DisAgg achieves a $1.37\times$ speedup over \textsc{OPA} after 30 iterations of FL training.

\subsection{SecAgg+/LightSecAgg Offline Processing}

\begin{figure}[!t]
    \centering
    \vspace{0.75em}
    \includegraphics[width=\columnwidth]{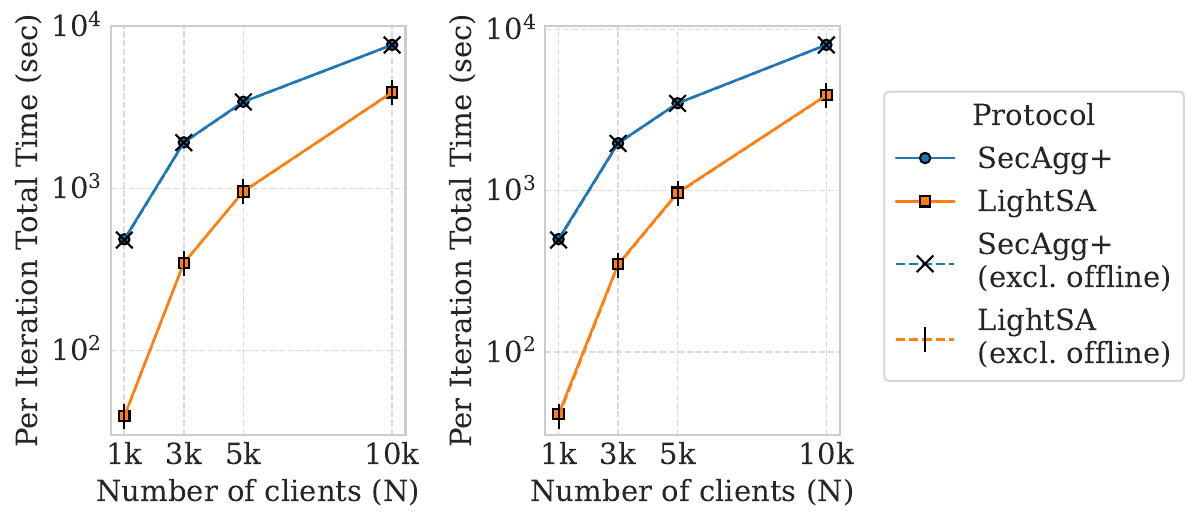}
    \vspace{-1.5em}
    \caption{Combined computation and communication time per FL iteration for \textsc{SecAgg+} and \textsc{LightSecAgg} (\textsc{LightSA}) for M=1k (left) and M=10k (right). Solid lines include mask computation time for \textsc{SecAgg+} and \textsc{LightSecAgg} while dashed lines exclude it. The difference in per iteration total time is negligible.}
    \label{fig:overall-compare-offline}
\end{figure}

\textsc{LightSecAgg} \cite{lightsecagg} introduces offline computation phases for clients -- allowing clients to compute mask shares in parallel to the FL training iteration. This effectively absorbs mask computation time into the total time spent for other phases. Figure \ref{fig:overall-compare-offline} simulates the effects of using offline phases for \textsc{SecAgg+} and \textsc{LightSecAgg} by including or excluding mask computation time from the per iteration total time. As seen, parallelizing mask creation has a negligible impact on the total time as bottlenecks lie in other phases, which can be resolved by using \textsc{DisAgg}.

\section{Artifact Appendix}

\subsection{Abstract}

This artifact provides the code and accompanying instructions required to reproduce the experimental results presented in the paper. The code includes Python implementations of the baseline methods compared with DisAgg and scripts to run experiments on the various datasets and models evaluated. A \texttt{README.md} file describes how to install the required dependencies, prepare the datasets, and execute the experiments. The instructions allow users to replicate the evaluation procedure and regenerate the performance metrics reported in the paper, including accuracy and execution time, subject to variations due to hardware differences.

\subsection{Artifact check-list (meta-information)}

{\small
\begin{itemize}
  \item {\bf Algorithm:} DisAgg
  \item {\bf Program:} Python 3.10 and various packages.
  \item {\bf Model:} CNN, TinyNet, EfficientNet, DistilBERT+LoRA
  \item {\bf Data set:} MNIST, CIFAR10, CIFAR100, CelebA, and SST2
  \item {\bf Run-time environment:} Linux 64-bit with Nvidia drivers, Python 3.10 and Pip installed.
  \item {\bf Hardware:} CPU with 10+ cores, GPU with at least 10 GB VRAM, at least 128 GB RAM, and 0.5 TB disk space
  \item {\bf Metrics:} Accuracy, Time, Communication Size
  \item {\bf Experiments:} see \texttt{README.md}
  \item {\bf How much disk space required (approximately):} 0.5 TB
  \item {\bf How much time is needed to prepare workflow (approximately):} 20mins
  \item {\bf How much time is needed to complete experiments (approximately):} Few minutes up to many hours, depending on hardware and experiment.
  \item {\bf Publicly available:} Yes
  \item {\bf Code licenses:} CC-BY-NC 4.0
  \item {\bf Data licenses:} See individual dataset URLs in \texttt{README.md}.
  \item {\bf Workflow framework used:} None
  \item {\bf Archived:} \href{https://zenodo.org/records/19462696}{DOI on Zenodo}
\end{itemize}
}

\subsection{Description}

\subsubsection{How delivered}

All instructions and code can be found using this publicly available URL: \url{https://github.com/SamsungLabs/mlsys26_disagg}. \texttt{README.md} inside the root directory contains instructions on how to install the required packages, prepare data and run experiments.

\subsubsection{Data sets}
The following datasets are used: MNIST, CIFAR-10, CIFAR-100, CelebA \& SST-2.
Instructions on how to download them can be found in the repo's \texttt{README.md}.

\subsection{Installation}

\texttt{README.md} contains instructions on how to install the required packages.

\subsection{Experiment workflow}

All experiments are configured in \texttt{src/constants.py}. To run a
protocol with experiment index \texttt{<i>}:
{\small
\begin{verbatim}
python -m disagg_test --exp_index=<ix>
python -m opa_test --exp_index=<i>
python -m light_secagg_test --exp_index=<i>
python -m secagg_plus_test --exp_index=<i>
\end{verbatim}
}

Each experiment index \texttt{<i>} corresponds to a specific result in a paper. All results are written to \texttt{outputs/}. Additional details on the experimental configuration and mapping between experiment indices and paper results can be found in the \texttt{README.md}.

\subsection{Experiment customization}

Experiments can be customized by modifying the configuration parameters in the source code. Additional details on how to change parameters and define new experimental settings are provided in the \texttt{README.md}.

\subsection{Evaluation and expected result}

\texttt{README.md} contains details on what experiments to run to replicate paper results. Note that timings would naturally vary based on hardware. However, we envision the comparative conclusions will be the same.

\subsection{Methodology}

Submission, reviewing and badging methodology:

\begin{itemize}
  \item \url{http://cTuning.org/ae/submission-20190109.html}
  \item \url{http://cTuning.org/ae/reviewing-20190109.html}
  \item \url{https://www.acm.org/publications/policies/artifact-review-badging}
\end{itemize}

\end{document}